\documentclass[fleqn,usenatbib]{mnras}

\usepackage{newtxtext,newtxmath}
\usepackage{xspace}

\usepackage[T1]{fontenc}
\usepackage{ae,aecompl}
\usepackage{url}

\usepackage{graphicx}	
\usepackage{amsmath}	
\usepackage{float}

\usepackage[utf8]{inputenc}

\title[New data on the photon-ALP mixing from blazars]{Revisiting the evidences for spectral anomalies in distant blazars: new data on the photon-ALP mixing
}

\author[F. Cenedese, A. Franceschini and G. Galanti]{Francesco Cenedese,$^{1}$\thanks{E-mail:
francesco.cenedese@hotmail.com}
Alberto Franceschini$^{1}$\thanks{E-mail:
alberto.franceschini@unipd.it}
and Giorgio Galanti$^{2}$\thanks{E-mail:
gam.galanti@gmail.com}\\
$^{1}$Dipartimento di Fisica e Astronomia, Universita' degli Studi di Padova, vicolo dell'Osservatorio 3, I -- 35122, Padova, Italy\\
$^{2}$INAF, Istituto di Astrofisica Spaziale e Fisica Cosmica di Milano, Via A. Corti 12, I -- 20133 Milano, Italy\\
}

\date{Received:~\today; Accepted:~ }

\begin{document}
\label{firstpage}
\pagerange{\pageref{firstpage}--\pageref{lastpage}}
\maketitle

\begin{abstract}
We re-examine possible dependencies on redshift of the spectral parameters of blazars observed at very-high energies (VHEs) with Imaging Atmospheric Cherenkov telescopes (IACTs). This is relevant to assess potential effects with the source distance of the photon to axion-like particle (ALP) mixing, that would deeply affect the propagation of VHE photons across the Universe.
We focus our spectral analysis on 38 BL Lac objects (32 high-peaked and 6 intermediate-peaked) up to redshift $z\simeq 0.5$, and a small sample of 5 Flat Spectrum Radio Quasars up to $z=1$ treated independently to increase the redshift baseline. 
The 78 independent spectra of these sources are first of all carefully corrected for the gamma-gamma interaction with photons of the Extragalactic Background Light, that are responsible for the major redshift-dependent opacity effect. Then, the corrected spectra are fitted with simple power-laws to infer the intrinsic spectral indices $\Gamma_{\rm em}$ at VHE, to test the assumption that such spectral properties are set by the local rather than the global cosmological environment.
We find some systematic anti-correlations with redshift of $\Gamma_{\rm em}$ that might indicate, although with low-significance, a {\it spectral anomaly} potentially requiring a revision of the photon propagation process.
More conclusive tests with higher statistical significance will require the observational improvements offered by the forthcoming new generation of Cherenkov arrays (CTA, ASTRI, LHAASO).

\end{abstract}


\section{Introduction}  
\label{sec:intro}

Gamma-ray astronomy plays a crucial role in the exploration of the most extreme and the most violent non-thermal phenomena in the Universe. This field has a great potential for discovery about many open questions of modern astrophysics and cosmology, such as the origin of galactic and extragalactic cosmic rays, particle acceleration and radiation processes under extreme astrophysical conditions. In addition, astrophysics at Very High Energies (VHE, that is for photons with energy exceeding $10^{11}$ eV) also offers tools for testing fundamental physics, like the search for dark matter, and tests of the Standard Model, possibly looking for evidences of new physics. These tests take advantage by the extreme energies of such photons, not reproducible in the laboratory. Observations of cosmic sources at VHE energies are possible with ground-based Imaging Atmospheric Cherenkov Telescopes \citep[e.g.][]{2004vhec.book.....A} and water-Cherenkov arrays, particularly LHAASO \citep[][]{2019ChAA..43..457C}.

Typically, these tests are based on energy-dependent effects emerging in the propagation of such high-energy photons during their travel across the Universe. In such a way, the universal validity of the Lorentz Invariance for example, or its violation (LIV, e.g. \citealt{Amelino1998}), can be tested, as well as the existence of exotic low-mass non-baryonic particles known as axion-like particles (ALPs, e.g. \citealt[][]{2011PhRvD..84j5030D}). LIV is a prediction of possible deviations from the Standard Model, in the framework of alternative theories of gravity and quantum gravity \citep{2008RPPh...71i6901A, Amelino1998}. Several studies about LIV consequences are present in the literature (see e.g. \citealt{Kifune1999,Stecker2001,JacobPiran2008,Fairbairn2014,TavecchioLIV,CTAfund}). The low-mass pseudo-scalar neutral ALPs are a generalization of the original axion (see e.g. \citealt{axion1,axion2,axion3}), they are invariably predicted by super-string theories (for a review see e.g. \citealt{ALPrev1,ALPrev2}) and represent a potential candidate for the cosmological Dark Matter \citep{preskill,abbott,dine,arias2012}. ALPs produce many consequences especially in high-energy astrophysics, such as the modification of the transparency of the medium crossed by the beam (e.g. \citealt{DRM2007,serpico,sanchezConde,2011PhRvD..84j5030D,2017PhRvD..96e1701K,grExt}), alterations in the observed spectra (e.g. \citealt{fermi2016,gtre2019,gtl2020,CTAfund}), and the modification of the polarization state of photons (e.g. \citealt{bassan,day,Galanti2022a,Galanti2022b,GRT2022}).\\
The advantage of VHE astrophysical tests, in addition to the high energies involved, is that, even if the interaction probability is exceedingly small, the enormous distances make such effects potentially well measurable.

As it is well known, however, the first order process affecting the propagation of VHE photons across space-time is their interaction with the low-energy Extragalactic Background Light (EBL), via the photon-photon interaction \citep[see for a review][]{2021Univ....7..146F}. Because of its strong dependence on energy, this produces an exponential cutoff in the spectra $\propto e^{-\tau_{\gamma\gamma}}$, where $\tau_{\gamma\gamma}$ is the optical depth to photon-photon interaction, increasing linearly with the proper distance to the source. Once this latter is known, $\tau_{\gamma\gamma}$ can be calculated based on models of the EBL \citep[e.g.][among others]{2008A&A...487..837F,2011MNRAS.410.2556D,2012MNRAS.422.3189G}, and the observed spectra of the gamma-ray sources can be corrected to obtain the emitted spectra at the source.

Once this correction is performed, the emitted VHE spectra can be assumed to be essentially independent of the large-scale environment, and particularly of the cosmic time, because the VHE emission can only be reasonably ruled by local physics inside the inner parsec-scale nucleus.

\begin{figure*}
\includegraphics[width=0.75\textwidth,height=0.5\textwidth]{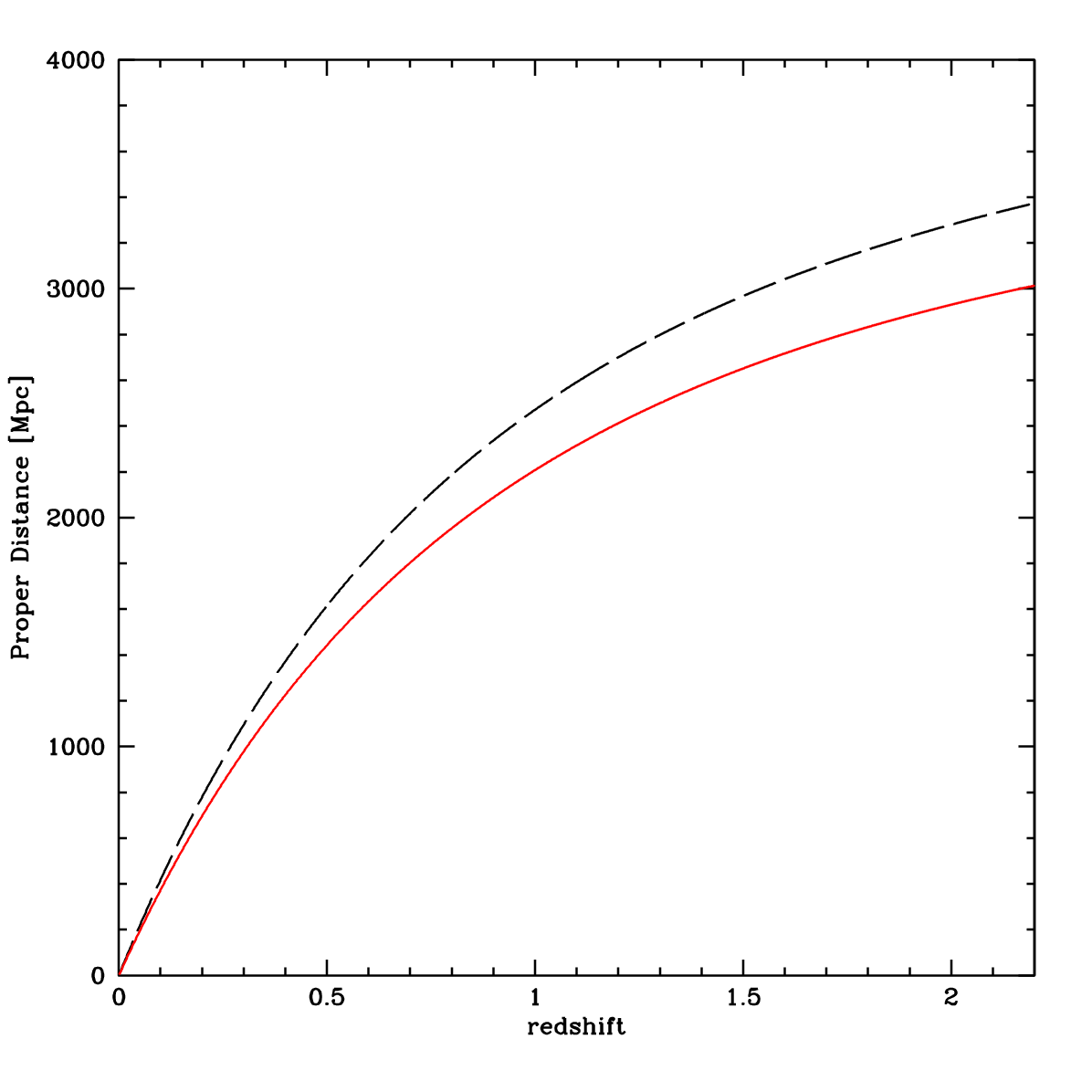}
\caption{ The graph quantifies the proper distance as a function of redshift, for two values of the Hubble constant: $H_0=67\ \rm Km/s/Mpc$ and $H_0=75\ \rm Km/s/Mpc$ corresponding to the black dashed and red continuous curves, respectively.
}
\label{fig:fig1}
\end{figure*}

Based on these assumptions, various authors have proposed to test for anomalies in the blazar spectra at high redshifts, that might be indicative of non-standard effects in the photon propagation. One of these is the previously mentioned possible existence of axions, or ALPs, whose expected behavior of mixing with photons would have potentially important observational consequences in terms of a reduced photon-photon opacity. 
Indeed, considering that ALPs are predicted to interact with two photons or with a photon and a static electromagnetic ${\bf E}$ and/or ${\bf B}$ field,
\cite{2009MNRAS.394L..21D} and \cite{2011PhRvD..84j5030D} suggested that in the presence of an environmental magnetic field, high energy photons and ALPs would oscillate, like it is the case for solar neutrinos: a VHE photon emitted by a gamma-ray source, by interacting with an intergalactic $\bf B$ field, would transform into an ALP, and the latter be reconverted in a photon after a subsequent interaction with another $\bf B$ field. Since during the ALP phase there is no interaction with the background photons and no pair production, this would overall reduce the photon-photon opacity.
Observations of VHE distant sources can then offer a potential to constrain the existence and behaviour of ALPs, by a correlation analysis of their gamma-ray spectral indices $\Gamma_{\rm em}(z)$ and their redshift $z$, since the effect is clearly dependent on the source distance. 
This path has been recently pursued by \cite {2020MNRAS.493.1553G} based on a limited sample of BL Lac objects selected with tight boundaries: to have secure redshifts $z< 0.6$, and be observed during flaring states.
From their analysis they find a statistical correlation between the EBL-corrected $\Gamma_{\rm em}(z)$ and $z$, with an hardening of the spectra with redshift that would conflict with expectations of conventional physics. Their proposed solution is a hint for ALP-to-photon mixing as responsible for reducing the optical depth $\tau_{\gamma\gamma}$, such that a conventional purely-EBL correction would erroneously over-correct the spectral data.

This result is relevant for various reasons.
The first obvious one is that, whenever confirmed, this would be a first astrophysical evidence that the standard model of fundamental interactions has to be improved more or less radically.
The second reason being that our treatment of the cosmic opacity to photon-photon interactions and pair-production has to be deeply revised, particularly in view of the next generation of IACT and water-Cherenkov arrays, with their much improved sensitivity at the highest energies. Should the Universe be so much more transparent above energies $E\geq 1$ TeV than normally expected, this would strongly affect our knowledge of high-energy emission processes in blazars.

For all these reasons we put under further scrutiny in this paper the issue of the possible {\it blazar spectral anomaly}, by considering a larger sample of BL Lacs currantly available, by considering also spectral data for blazars in the low-activity states, and by performing a rigorous statistical investigation. At variance with \cite {2020MNRAS.493.1553G}, we have preferred here to proceed to a complete re-fitting of all our collected spectral data, that simplifies and strengthens the analysis.
 
Also, because the departure of the ALP-dominated effect is dependent on cosmic proper distance to the source, we consider in addition a small sample of Flat-Spectrum Radio Quasars (FSRQs) observed at VHE energies up to $z=1$, allowing us to substantially expand the redshift baseline.
How much the proper distance to the source depends on $z$ is illustrated in Fig. \ref{fig:fig1}. We see that, from $z=0.4$ to $z=1$, the integrated  distance increases by a factor of about 2.

The paper is structured as follows.
Section 2 is dedicated to a review of the blazar sample used for our analysis, including BL Lac objects and FSRQs.
Section 3 describes our adopted model for the EBL photon density, and the procedure for evaluating the attenuation of gamma-rays due to pair production. 
Our results are reported in Section 4 based on the pure EBL pair-production correction and in Section 5 for the ALP corrections.
Implications of these results are briefly reviewed in Section 6.

In the present paper we adopt the following cosmology: $H_0=70\ {\rm km/s/Mpc}$ and $\Omega_\Lambda=0.7$, $\Omega_m=0.3$.

\begin{figure*}
\includegraphics[width=0.60\textwidth]{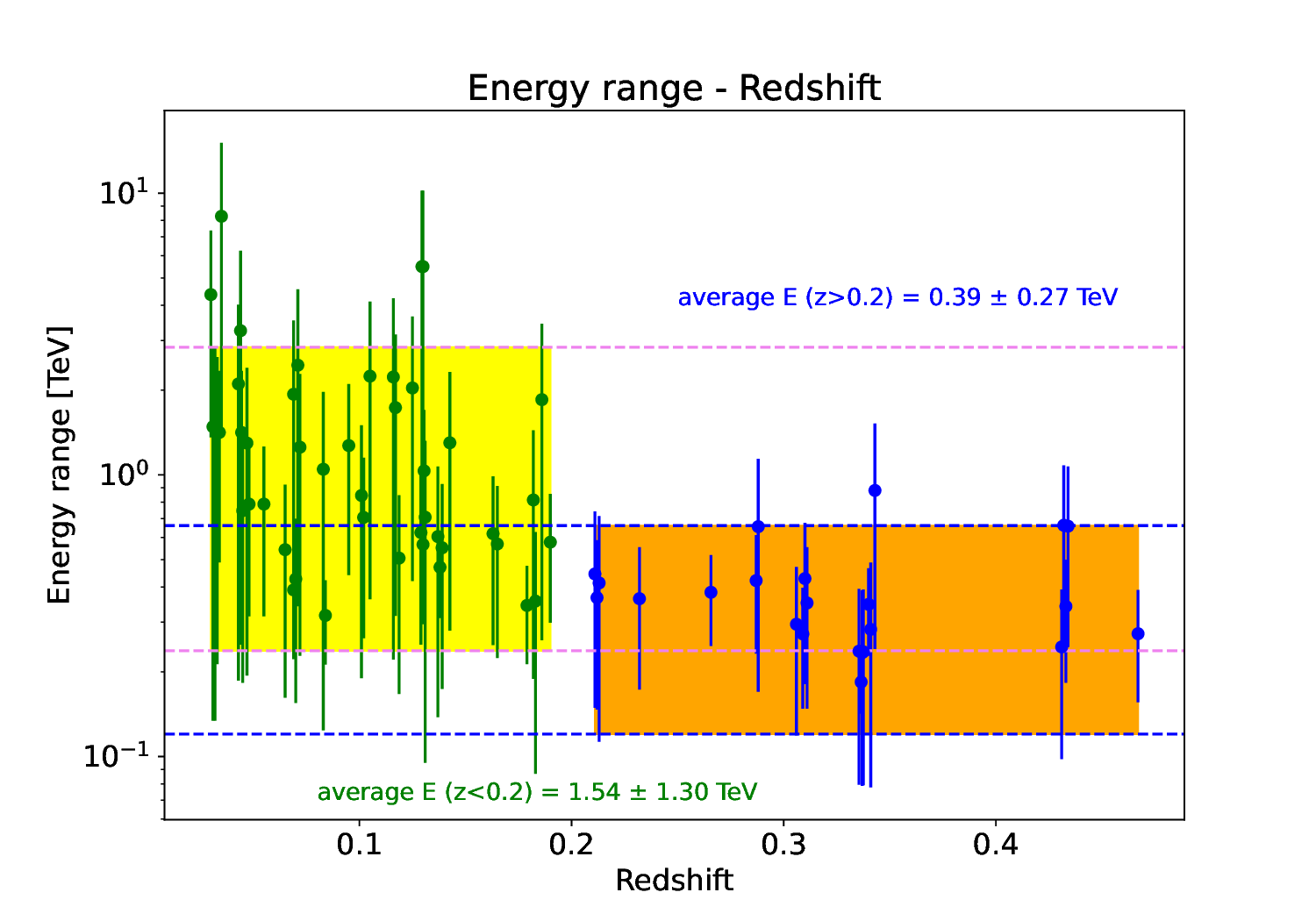}
\caption{A sketch of the spectral energy range of our sample of BL Lac objects. We highlight the two different coverages for the $z<0.2$ (yellow) and $z>0.2$ (orange) sub-samples.
Note that in case of overlap the points in the plot are slightly shifted for clarity.
}
\label{fig:fig2}
\end{figure*}

\begin{figure*}
\includegraphics[width=0.6\textwidth]{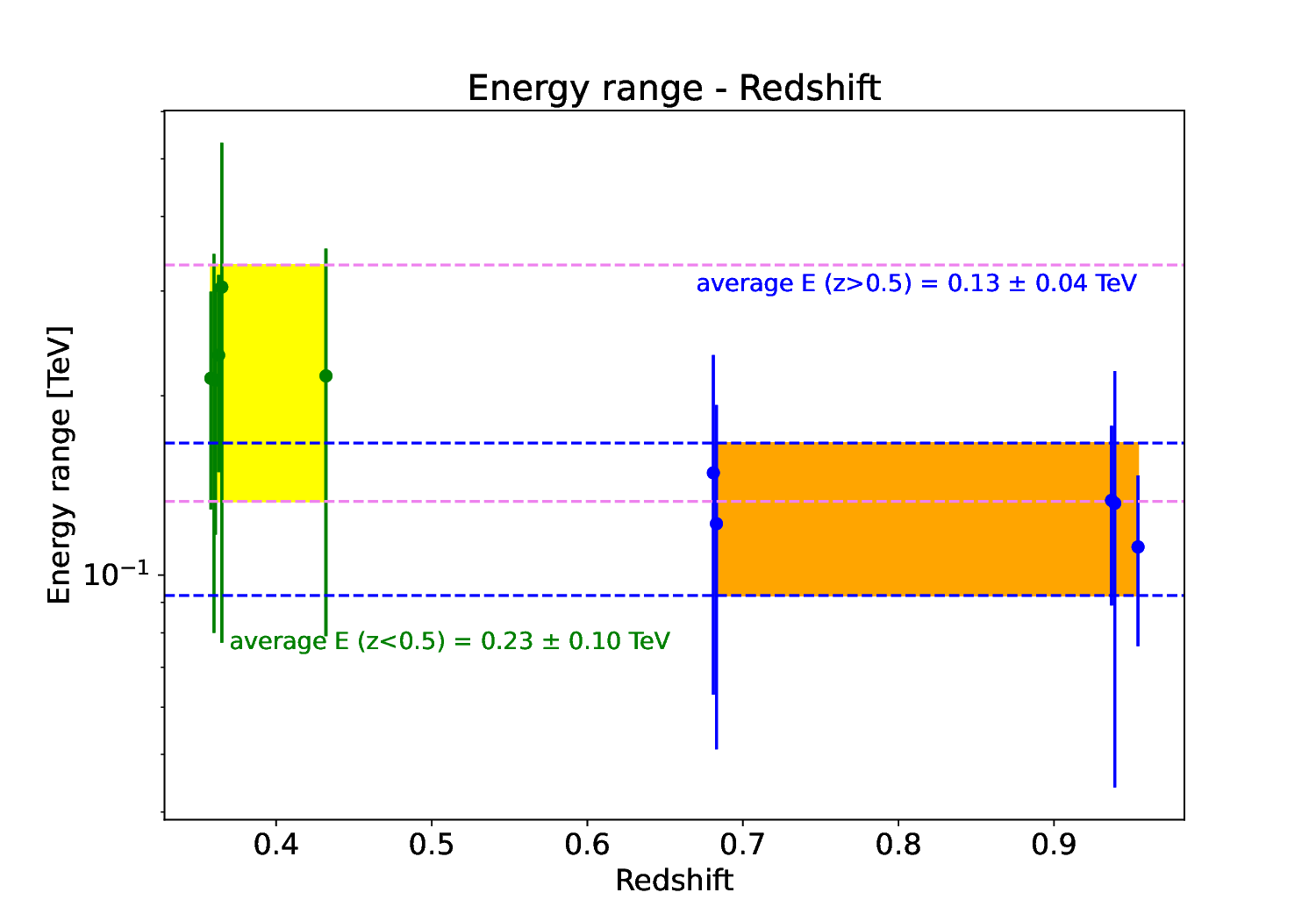}
\caption{A sketch of the spectral energy range of our sample of FSRQs, with the low-$z$ and high-$z$ samples highlighted. See also caption of Fig. \ref{fig:fig2}.
}
\label{fig:fig3}
\end{figure*}

\begin{table*}
\begin{center}
\caption{Sample blazars with $z$ < 0.15.}\label{tab:table1}
\begin{tabular}{llccccc} 
\hline
Source           & $z$ & Detector & $\Delta E_{0}(z)$   & $\Gamma_{\rm obs}$ & K$_{\rm obs}$ & References\\
(spectral type)                 &   &        & [TeV]  &            &   [$\rm cm^{-2} \, s^{-1} \, TeV^{-1}$] & \\
\hline 
1RXS J101015.9-311909 & 0.143  & HESS  & 0.280-2.32 &  3.03$\pm$0.48  & (7.77 $\pm$ 2.83) $\times 10^{-12}$ & \cite{2012AA...542A..94H} 
\\
1ES 0806+524 & 0.138  & VERITAS  & 0.310-0.630 &  3.63$\pm$0.20  & (1.95 $\pm$ 0.17) $\times 10^{-11}$ & \cite{2009ApJ...690L.126A} \\
 &   & MAGIC high-state  & 0.174-0.928 &  2.91$\pm$0.18  & (5.55 $\pm$ 0.54) $\times 10^{-11}$ & \cite{2015MNRAS.451..739A} \\
 &   & MAGIC low-state  & 0.138-1.070 &  2.77$\pm$0.25  & (1.58 $\pm$ 0.24) $\times 10^{-11}$ & \cite{2015MNRAS.451..739A} \\
1ES 1215+303 & 0.13  & VERITAS  & 0.295-0.837 &  3.70$\pm$0.31  & (2.23 $\pm$ 0.37) $\times 10^{-11}$ & \cite{2013ApJ...779...92A} \\
 &   & MAGIC & 0.095-1.320 &  2.97$\pm$0.16  & (2.17 $\pm$ 0.28) $\times 10^{-11}$ & \cite{2012AA...544A.142A} 
 \\
H 1426+428  & 0.129  & CAT  & 0.250-0.998 &  3.73$\pm$0.20  & (3.47 $\pm$ 0.50) $\times 10^{-10}$ & \cite{2011APh....34..674M} \\
 &   & HEGRA 2000  & 0.774-10.200 &  2.74$\pm$0.36  & (8.01 $\pm$ 6.54) $\times 10^{-11}$ & \cite{2011APh....34..674M} \\
 &   & HEGRA 2002  & 0.778-10.200 &  2.02$\pm$0.13  & (5.61 $\pm$ 1.85) $\times 10^{-12}$ & \cite{2011APh....34..674M} \\
  &   & Whipple  & 0.367-1.700 &  3.46$\pm$0.10  & (2.98 $\pm$ 0.27) $\times 10^{-10}$ & \cite{2011APh....34..674M} \\
RGB J0710+591  & 0.125  & VERITAS  & 0.420-3.650 &  2.68$\pm$0.19  & (2.48 $\pm$ 0.62) $\times 10^{-11}$ & \cite{2010ApJ...715L..49A} \\
B3 2247+381  & 0.1187  & MAGIC  & 0.167-0.846 &  3.29$\pm$0.13  & (1.37 $\pm$ 0.10) $\times 10^{-11}$ & \cite{2012AA...539A.118A} \\
PKS 2155-304  & 0.116  & HESS  & 0.221-4.230 &  3.53$\pm$0.08  & (1.20 $\pm$ 0.08) $\times 10^{-10}$ & \cite{2010AA...520A..83H} \\
 &  & MAGIC  & 0.316-3.150 &  3.26$\pm$0.12  & (8.76 $\pm$ 1.00) $\times 10^{-10}$ & \cite{2012AA...544A..75A} \\
1ES 1312-423  & 0.105  & HESS  & 0.362-4.120 &  2.67$\pm$0.15  & (5.89 $\pm$ 1.26) $\times 10^{-12}$ & \cite{2013MNRAS.434.1889H} \\
W Comae (IBL) & 0.102  & VERITAS & 0.263-1.150 &  3.74$\pm$0.37  & (6.02 $\pm$ 0.97) $\times 10^{-11}$ & \cite{2008ApJ...684L..73A} \\
 &   & VERITAS II  & 0.190-1.500 &  3.68$\pm$0.19  & (2.00 $\pm$ 0.25) $\times 10^{-10}$ & \cite{2009ApJ...707..612A} \\
SHBL J001355.9-185406 & 0.095  & HESS  & 0.441-2.100 &  3.45$\pm$0.05  & (7.04 $\pm$ 0.29) $\times 10^{-12}$ & \cite{2013AA...554A..72H} \\
1ES 1741+196 & 0.084  & VERITAS  & 0.212-0.422 &  2.72$\pm$0.14  & (9.30 $\pm$ 0.34) $\times 10^{-12}$ & \cite{2016MNRAS.459.2550A} \\
 &  & MAGIC  & 0.124-1.970 &  2.38$\pm$0.14  & (4.88 $\pm$ 0.77) $\times 10^{-12}$ & \cite{2017MNRAS.468.1534A} \\
PKS 2005-489  & 0.071  & HESS  & 0.342-4.560 &  3.20$\pm$0.17  & (3.45 $\pm$ 0.62) $\times 10^{-11}$ & \cite{2010AA...511A..52H} \\
 &  & HESS II  & 0.228-2.280 &  3.98$\pm$0.33  & (2.36 $\pm$ 0.43) $\times 10^{-11}$ & \cite{2005AA...436L..17A} \\
BL Lacertae (IBL) & 0.069  & VERITAS  & 0.222-0.559 &  3.75$\pm$0.50  & (5.63 $\pm$ 0.83) $\times 10^{-10}$ & \cite{2013ApJ...762...92A} \\
 &   & MAGIC  & 0.155-0.699 &  3.63$\pm$0.16  & (1.88 $\pm$ 0.14) $\times 10^{-11}$ & \cite{2007ApJ...666L..17A} \\
PKS 0548-322  & 0.069  & HESS  & 0.337-3.530 &  2.42$\pm$0.35  & (8.30 $\pm$ 3.66) $\times 10^{-12}$ & \cite{2010AA...521A..69A} \\
PKS 1440-389  & 0.065  & HESS  & 0.162-0.923 &  3.39$\pm$0.27  & (2.23 $\pm$ 0.34) $\times 10^{-11}$ & \cite{2020MNRAS.494.5590A} \\
1ES 1727+502  & 0.055  & VERITAS  & 0.315-1.260 &  2.18$\pm$0.26  & (3.65 $\pm$ 0.83) $\times 10^{-11}$ & \cite{2015ApJ...808..110A} \\
1ES 1959+650  & 0.048  & MAGIC  & 0.193-1.530 &  2.80$\pm$0.07  & (9.11 $\pm$ 0.52) $\times 10^{-11}$ & \cite{2006ApJ...639..761A} \\
 &   & MAGIC II  & 0.194-2.40 &  2.54$\pm$0.11  & (5.69 $\pm$ 0.66) $\times 10^{-11}$ & \cite{2008ApJ...679.1029T} \\
 Markarian 180  & 0.045  & MAGIC  & 0.183-1.310 &  3.18$\pm$0.10  & (4.18 $\pm$ 0.34) $\times 10^{-11}$ & \cite{2006ApJ...648L.105A} \\
 1ES 2344+514 & 0.044  & VERITAS  & 0.250-6.250 &  2.46$\pm$0.06  & (4.03 $\pm$ 0.30) $\times 10^{-11}$ & \cite{2017MNRAS.471.2117A} \\
 &   & MAGIC  & 0.186-4.020 &  2.99$\pm$0.18  & (6.53 $\pm$ 0.88) $\times 10^{-11}$ & \cite{2007ApJ...662..892A} \\
 &   & VERITAS II & 0.489-2.340 &  2.44$\pm$0.14  & (3.22 $\pm$ 0.49) $\times 10^{-10}$ & \cite{2011ApJ...738..169A} \\
Markarian 501 & 0.034  & VERITAS  & 0.251-2.510 &  2.72$\pm$0.21  & (1.58 $\pm$ 0.28) $\times 10^{-10}$ & \cite{2011ApJ...729....2A} \\
 &  & MAGIC  & 0.213-2.620 &  2.62$\pm$0.14  & (1.95 $\pm$ 0.22) $\times 10^{-10}$ & \cite{2011ApJ...729....2A} \\
  &   &  HAWC & 1.460-15.100 &  2.94$\pm$0.05  & (4.92 $\pm$ 0.57) $\times 10^{-10}$ & \cite{2019ICRC...36..654C} \\
Markarian 421  & 0.031  & MAGIC  & 0.134-2.830 &  2.87$\pm$0.07  & (4.22 $\pm$ 0.36) $\times 10^{-10}$ & \cite{2007ApJ...663..125A} \\
 &   & MAGIC  & 0.134-2.856 &  2.80$\pm$0.08  & (4.55 $\pm$ 0.26) $\times 10^{-10}$ & \cite{2007ApJ...663..125A} \\
 &   & HAWC  & 1.360-7.370 &  3.23$\pm$0.12  & (1.85 $\pm$ 0.52) $\times 10^{-9}$ & \cite{2019ICRC...36..654C} \\
\hline
\end{tabular}
\end{center}
\raggedright
\footnotesize 
\texttt{Col.1}: Object name reported in the TeVCat catalog; all sources are classified as HBL, except those otherwise indicated; \texttt{Col.2}: Redshift of the source (as reported in TeVCat) ; \texttt{Col.3}: Detector used for observation. \texttt{Col.4}: Observational data energy range; \texttt{Col.5-6}: Spectral observed index and normalization constant at photon energy $\epsilon_0=300\ \rm GeV$ (see Eq. \ref{S}). The errors indicated in \texttt{Col.5} are only the statistical errors. Statistical and systematic errors are added in quadrature to produce the total error reported on the plot spectral slope. Systematic errors are taken to be 0.1 for H.E.S.S., 0.15 for VERITAS, 0.2 for MAGIC and 0.3 for CAT, HEGRA and Whipple.  
\end{table*}

\begin{table*}
\begin{center}
\caption{Sample blazars with $z$ > 0.15.}\label{tab:table2}
\begin{tabular}{llccccc} 
\hline
Source           & $z$ & Detector & $\Delta E_{0}$(z)   & $\Gamma_{\rm obs}$ & K$_{\rm obs}$ & References\\
(spectral type)                 &   &        & [TeV]  &            &   [$\rm cm^{-2} \, s^{-1} \, TeV^{-1}$] & \\
\hline                                                  
  
1ES 0033+595 
 & 0.467 & MAGIC  & 0.156-0.391 &  3.84$\pm$0.01  & (9.73 $\pm$ 0.01) $\times 10^{-12}$ & \cite{2015MNRAS.446..217A}  \\
 PG 1553+113 & 0.433  & VERITAS  & 0.183-0.500 & 4.45$\pm$0.15  & (5.23 $\pm$ 0.24) $\times 10^{-11}$ & \cite{2015ApJ...799....7A}\\
 &   & HESS & 0.245-1.080 &  4.01$\pm$0.35 & (5.68 $\pm$ 0.75) $\times 10^{-11}$ & \cite{2008AA...477..481A}
 \\
 &   & HESS 2005+2006 & 0.245-1.070 &  4.44$\pm$0.50  & (4.60 $\pm$ 0.61) $\times 10^{-11}$ & \cite{2008AA...477..481A}
 \\ 
 &   & MAGIC & 0.098-0.392 &  4.06$\pm$0.18  & (3.73 $\pm$ 0.50) $\times 10^{-11}$ & \cite{2007ApJ...654L.119A} \\
PKS 0447-439 & 0.343  & HESS & 0.241-1.520 &  3.85$\pm$0.48  & (3.48 $\pm$ 0.93) $\times 10^{-11}$ & \cite{2013AA...552A.118H}
\\ 
3C 66A (IBL) & 0.34  & VERITAS  & 0.228-0.466 &  4.08$\pm$0.28  & (4.12 $\pm$ 0.28) $\times 10^{-11}$ & \cite{2009ApJ...693L.104A}\\
&  & MAGIC  & 0.078-0.488 &  3.44$\pm$0.16  & (2.04 $\pm$ 0.29) $\times 10^{-11}$ & \cite{2011ICRC....8...97K}\\
TXS 0506+056 (blazar) & 0.3365  & VERITAS  & 0.142-0.226 &  4.85$\pm$0.97  & (2.26 $\pm$ 1.35) $\times 10^{-12}$ & \cite{2018ApJ...861L..20A}\\ 
&  & MAGIC A & 0.079-0.392 &  3.77$\pm$0.08  & (1.87 $\pm$ 0.17) $\times 10^{-11}$ & \cite{2018ApJ...863L..10A}\\
 &  & MAGIC B & 0.080-0.394 &  3.49$\pm$0.51  & (2.62 $\pm$ 1.45) $\times 10^{-11}$ & \cite{2018ApJ...863L..10A}\\
 &  & MAGIC C & 0.079-0.389 &  3.79$\pm$0.38  & (6.01 $\pm$ 2.49) $\times 10^{-12}$ & \cite{2018ApJ...863L..10A}\\
S5 0716+714 (IBL) & 0.31  & MAGIC 2008  & 0.181-0.676 &  3.38$\pm$0.48  & (1.40 $\pm$ 0.35) $\times 10^{-10}$ & \cite{2009ApJ...704L.129A}\\
&  & MAGIC 2015 A & 0.148-0.554 &  4.07$\pm$0.07  & (6.77 $\pm$ 0.24) $\times 10^{-11}$ & \cite{2018AA...619A..45M}
\\
 &  & MAGIC 2015 B & 0.148-0.397 &  4.60$\pm$0.14  & (4.78 $\pm$ 0.36) $\times 10^{-11}$ & \cite{2018AA...619A..45M}
 \\
OJ 287 (BL Lac) & 0.306  & VERITAS  & 0.119-0.471 &  3.49$\pm$0.13  & (6.88 $\pm$ 0.47) $\times 10^{-12}$ & \cite{2017arXiv170802160O}\\
1ES 0414+009 & 0.287  & VERITAS  & 0.232-0.611 &  3.40$\pm$0.44  & (1.67 $\pm$ 0.25) $\times 10^{-11}$ & \cite{2012ApJ...755..118A}\\
 &  & HESS & 0.170-1.14 &  3.41$\pm$0.16  & (5.78 $\pm$ 0.41) $\times 10^{-12}$ & \cite{2012AA...538A.103H}
 \\
PKS 0301-243 & 0.2657  & HESS  & 0.247-0.519 &  4.42$\pm$0.65  & (9.38 $\pm$ 1.57) $\times 10^{-12}$ & \cite{2013AA...559A.136H}
\\
1RXS J023832.6-311658 & 0.232  & HESS  & 0.173-0.554 &  3.49$\pm$0.73  & (6.03 $\pm$ 1.62) $\times 10^{-12}$ & \cite{2017ICRC...35..645G}\\
1ES 1011+496 & 0.212  & MAGIC  & 0.147-0.586 &  4.13$\pm$0.47  & (3.66 $\pm$ 1.04) $\times 10^{-11}$ & \cite{2007ApJ...667L..21A}\\
 & & Ahnen  & 0.149-0.741 &  3.28$\pm$0.18  & (4.80 $\pm$ 0.41) $\times 10^{-11}$ & \cite{2016MNRAS.459.2286A}\\
 & & Aleksic  & 0.113-0.714 &  3.68$\pm$0.17  & (3.00 $\pm$ 0.32) $\times 10^{-11}$ & \cite{2016AA...591A..10A}
 \\
RBS 0413 & 0.19  & VERITAS  & 0.299-0.855 &  3.20$\pm$0.18  & (1.40 $\pm$ 0.14) $\times 10^{-11}$ & \cite{2012ApJ...750...94A}\\
1ES 1101-232 & 0.186  & HESS  & 0.259-3.44 &  2.95$\pm$0.18  & (1.96 $\pm$ 0.34) $\times 10^{-11}$ & \cite{2007AA...470..475A}
\\
1ES 1218+304 & 0.182  & VERITAS  & 0.189-1.44 &  3.14$\pm$0.22  & (3.63 $\pm$ 0.57) $\times 10^{-11}$ & \cite{2009ApJ...695.1370A}\\
 &  & MAGIC  & 0.087-0.626 &  3.03$\pm$0.21  & (4.65 $\pm$ 0.67) $\times 10^{-11}$ & \cite{2006ApJ...642L.119A}\\
RX J0648.7+1516 & 0.179  & VERITAS  & 0.213-0.475 &  4.36$\pm$0.42  & (2.28 $\pm$ 0.27) $\times 10^{-11}$ & \cite{2011ApJ...742..127A}\\
H 2356-309 & 0.165  & HESS  & 0.224-0.913 &  3.02$\pm$0.25  & (1.20 $\pm$ 0.14) $\times 10^{-11}$ & \cite{2006AA...455..461A}
\\
1ES 1440+122 & 0.163  & VERITAS  & 0.249-0.988 &  3.10$\pm$0.46  & (6.98 $\pm$ 1.73) $\times 10^{-12}$ & \cite{2016MNRAS.461..202A}\\
\hline
\end{tabular}
\end{center}
\raggedright
\footnotesize 
\texttt{Col.1}: Object name reported in the TeVCat catalog; all sources are classified as HBL, except those otherwise indicated; \texttt{Col.2}: Redshift of the source (as reported in TeVCat) ; \texttt{Col.3}: Detector used for observation. \texttt{Col.4}: Observational data energy range; \texttt{Col.5-6}: Spectral observed index and normalization constant at photon energy $\epsilon_0=300\ \rm GeV$ (see Eq. \ref{S}). The errors indicated in \texttt{Col.5} are only the statistical errors. Statistical and systematic errors are added in quadrature to produce the total error reported on the plot spectral slope. Systematic errors are taken to be 0.1 for H.E.S.S., 0.15 for VERITAS, 0.2 for MAGIC. 
\end{table*}

\begin{table*}
\begin{center}
\caption{Sample FSRQs.}\label{tab:table3}
\begin{tabular}{llccccccc} 
\hline
Source           & $z$ & Detector & $\Delta E_{0}$(z)   & $\Gamma_{\rm obs}$ & K$_{\rm obs}$ & References\\
                 &   &        & [TeV]  &            &   [$\rm cm^{-2} \, s^{-1} \, TeV^{-1}$] &\\
\hline                                                  
S3 0218+35 & 0.954  & MAGIC  & 0.076-0.147 & 3.93$\pm$0.45  & (2.92 $\pm$ 1.47) $\times 10^{-11}$ & \cite{2016AA...595A..98A}\\
PKS 1441+25 & 0.939  & MAGIC & 0.044-0.220 &  4.65$\pm$0.22 & (6.08 $\pm$ 1.88) $\times 10^{-12}$ & \cite{2015ApJ...815L..23A}\\
 &   & VERITAS & 0.089-0.178 &  5.00$\pm$0.73  & (3.12 $\pm$ 2.25) $\times 10^{-12}$ & \cite{2015ApJ...815L..22A} \\ 
 B2 1420+32 & 0.682  & MAGIC fase C & 0.051-0.193 &  4.09$\pm$0.19  & (2.70 $\pm$ 0.78) $\times 10^{-11}$ & \cite{2021AA...647A.163M} \\
 &  & MAGIC fase D & 0.063-0.234 &  4.17$\pm$0.09  & (1.05 $\pm$ 0.14) $\times 10^{-11}$ & \cite{2021AA...647A.163M} \\ 
4C +21.35 & 0.432  & MAGIC & 0.079-0.353 &  3.74$\pm$0.21  & (1.71 $\pm$ 0.32) $\times 10^{-10}$ & \cite{2011ApJ...730L...8A} \\ 
PKS 1510-089 & 0.361  & MAGIC 2012 & 0.117-0.308 &  3.63$\pm$0.08  & (1.23 $\pm$ 0.05) $\times 10^{-11}$ & \cite{2018AA...619A.159M} \\
&  & MAGIC 2012-2017 & 0.080-0.346 &  3.73$\pm$0.13  & (6.19 $\pm$ 0.81) $\times 10^{-12}$ & \cite{2018AA...619A.159M} \\
 &  & MAGIC 2015  & 0.129-0.299 &  4.50$\pm10^{-14}$  & 4.03 $\times 10^{-11}$ $\pm 10^{-25}$ & \cite{2018AA...619A.159M} \\ 
 &  & MAGIC 2016 & 0.077-0.532 &  4.34$\pm$0.19  & (1.76 $\pm$ 0.32) $\times 10^{-10}$ & \cite{2018AA...619A.159M} \\
\hline
\end{tabular}
\end{center}
\raggedright
\footnotesize 
\texttt{Col.1}: Object name reported in the TeVCat catalog; \texttt{Col.2}: Redshift of the source (as reported in TeVCat) ; \texttt{Col.3}: Detector used for observation. \texttt{Col.4}: Observational data energy range; \texttt{Col.5-6}: Spectral observed index and normalization constant at photon energy $\epsilon_0=300\ \rm GeV$ (see Eq. \ref{S}). The errors indicated in \texttt{Col.5} are only the statistical errors. Statistical and systematic errors are added in quadrature to produce the total error reported on the plot spectral slope. Systematic errors are taken to be 0.15 for VERITAS and 0.2 for MAGIC. 
\end{table*}

\begin{figure*}
\includegraphics[width=0.70\textwidth]{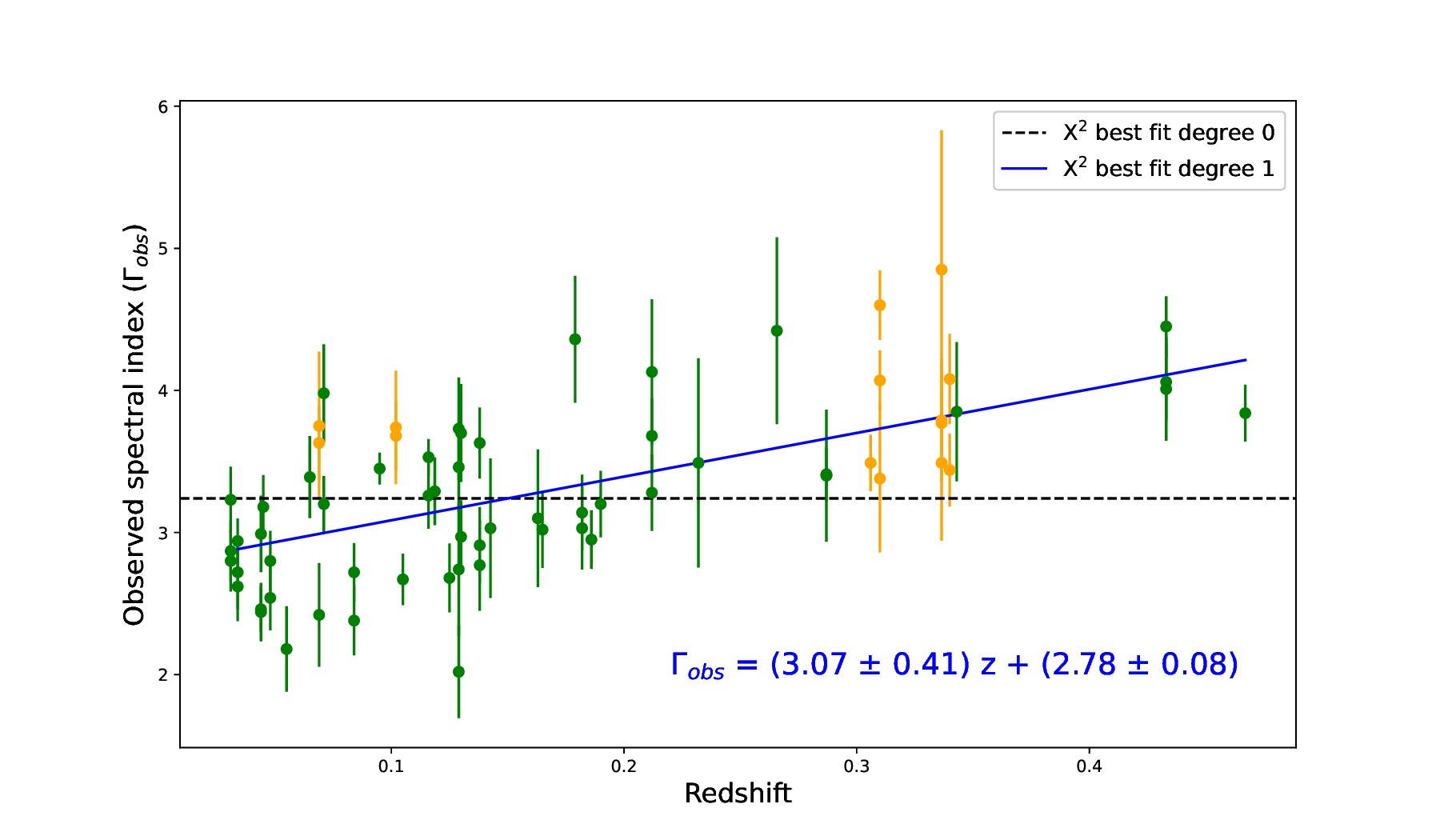}
\caption{ Plot of the observed spectral indices vs redshift for the BL Lac blazars, as derived from our spectral fits to the measured data.
The best-fit line was estimated with a chi squared test, obtaining $\chi^2_\nu=2.61$ due to the large scatter of the data. Green and yellow colors are HBL and IBL respectively.
}
\label{fig:fig4}
\end{figure*}

\section{The Dataset}
\label{sec:dataset}

Because blazars are by far the most numerous population of sources emitting at VHE, we use them for our spectral tests, also  taking advantage of their simple power-law spectra.
At variance with previous analyses, we use both BL Lacs and FSRQs, considering the latter useful to expand the redshift coverage. 
However, while both categories were analyzed in the same way, because BL Lacs and FSRQs have different astrophysical properties, we analyzed them separately.

For our study it was essential to know the source redshift, the observed spectrum and the energy range in which every blazar is observed. Our assumed spectral threshold is that corresponding to the classical VHE regime, i.e. including all photons with observed energy $E_{0}\geq100$ GeV. 
The basic information on the sources was obtained from the TeVCat\footnote{\url{http://tevcat.uchicago.edu}} reference catalogue \citep[][]{2008ICRC....3.1341W}. From this, for each source, the redshift and the publications reporting the observed spectra were obtained (typically from articles by the large IACT collaborations, HESS, MAGIC and VERITAS). 
If not present in TeVCat, we inferred this information from a dedicated search in the literature. All blazars with this information available were used for this study. Furthermore, if a source had multiply observed spectra, the analysis was done for all the individual spectra separately. 

\begin{figure*}
\includegraphics[width=0.70\textwidth,height=0.45\textwidth]{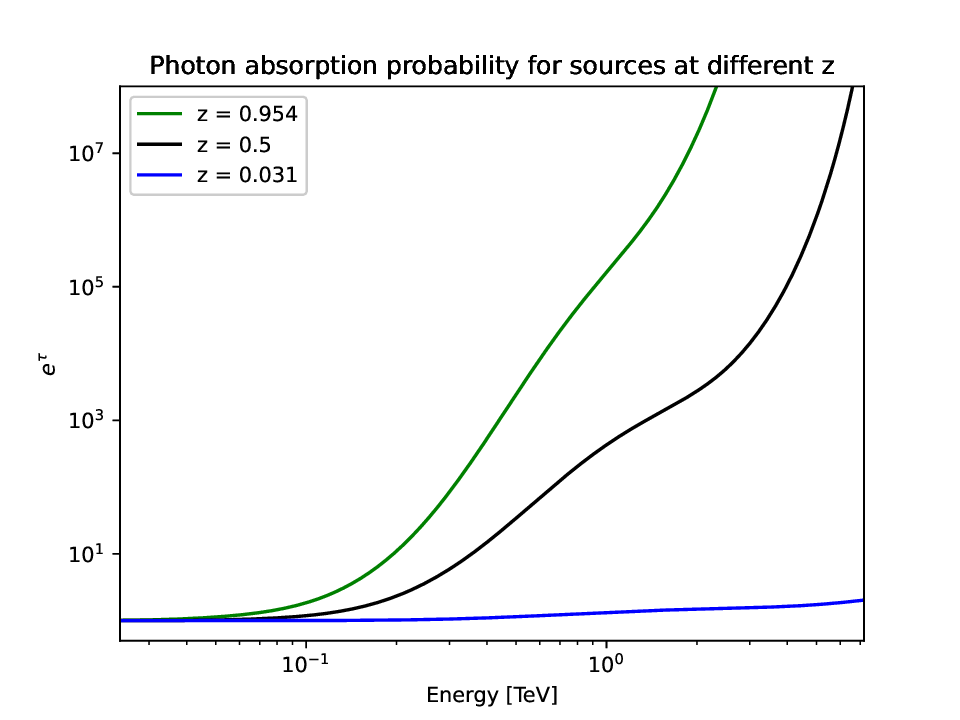}
\caption{ Extinction factors for gamma-ray sources at three different redshifts, due to the interaction with the EBL low-energy photons. The factors are plotted as a function of the gamma-ray photon energy. This strong dependence on energy produces an exponential cutoff in the VHE spectra of blazars.
}
\label{abs}
\end{figure*}

\begin{figure*}
\includegraphics[width=0.70\textwidth,height=0.45\textwidth]{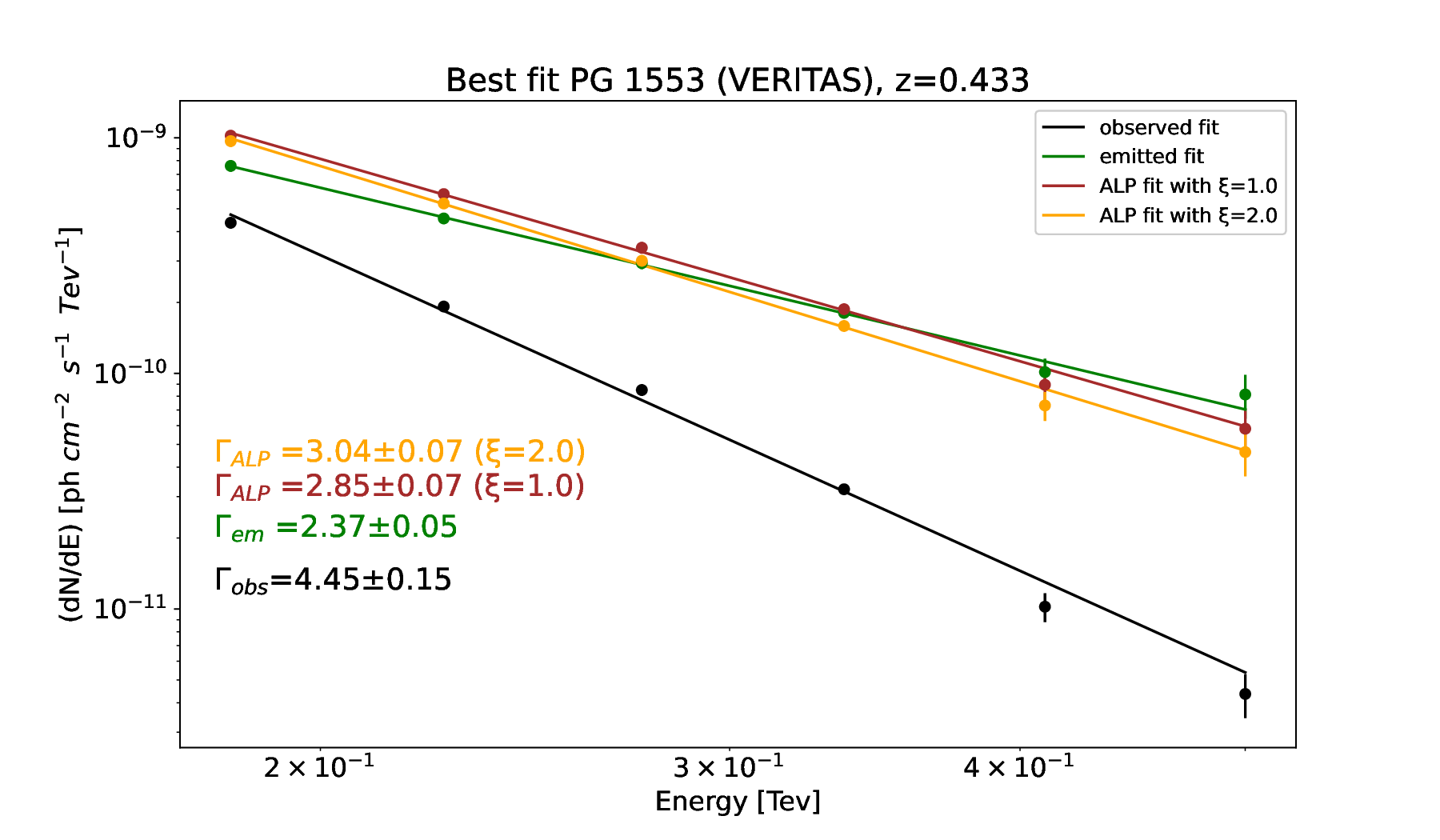}
\caption{ An example of photon-photon opacity corrections for the high-redshift BL Lac PG 1553+113 at $z=0.433$ observed by the VERITAS observatory (see Table \ref{tab:table2}). The lower power-law fit and data-points are the observed one, the upper green ones and best-fit are after EBL correction. Spectral data and lines denoted as ALP fits correspond to opacity corrections based on ALP models discussed in Sect. \ref{corrALP}.
}
\label{pg1553}
\end{figure*}

A sketch of the observed photon energy ranges of BL Lacs and FSRQs as a function of $z$ is shown in Figs. \ref{fig:fig2} and \ref{fig:fig3}, respectively. The ranges shown correspond to the maximum and minimum energy point of the observed spectrum, the data-points to their average value.   
The colored areas indicate $\sim 80\%$ of the energy ranges. Energy ranges for sources above and below $z = 0.2$ are highlighted. 
The average energy range for sources at $z < 0.2$ is obviously higher than that for sources at $z > 0.2$ due to the absorption of VHE photons by the EBL in the higher-$z$ sample. For the same reason the energy ranges for $z < 0.2$ are much wider. Similar considerations apply to FSRQs.

After obtaining from the literature the spectral data for all the sources, we have fitted them with power-law functions of the VHE photon energy $\epsilon$ and of the spectral index $\Gamma_{\rm obs}$
\begin{equation}
S_{\rm obs}(\epsilon,z)=K_{\rm obs} \left[\frac{\epsilon}{\epsilon_{0}}\right]^{-\Gamma_{\rm obs}} 
\label{S}
\end{equation}
where $K_{\rm obs}$ and $\epsilon_{0}=300\ \rm GeV$ are normalization constants.

\subsection{The BL Lac sample}

BL Lac objects are classified in the literature into three categories according to the frequency $\nu_{\rm sync}$ of the synchrotron peak: sources with $\nu_{\rm sync}<10^{14} \, \rm Hz$ are cataloged as low BL Lacs (LBLs), those with $10^{14}<\nu_{\rm sync}<10^{15} \, \rm Hz$ as intermediate BL Lacs (IBLs), and finally those with $\nu_{\rm sync}>10^{15} \, \rm Hz$ high BL Lacs (HBLs). In order to achieve a large but relatively homogeneous sample to work on, in this study it was decided to consider only the IBLs and HBLs. These objects represent 97\% of the BL Lacs observed at VHE so far.\\
The overall BL Lac sample analyzed in this paper consists of 38 sources, for which we collected altogether 69 spectra observed at VHE, in some cases spectra of the same source taken at different times and/or with different instruments (the "multi-spectrum" sources).
Of the 38 sources, 32 are catalogued in TeVCat as HBL (15 multi-spectrum), 4 as IBL (4 multi-spectrum), 1 simply as "Blazar" and 1 as "BL Lac" (multi-spectrum). The last two sources were included among the IBLs in the analysis in order not to contaminate the results of the HBL sources. However, a note was included to highlight the results of the IBLs analysis without considering the two extra sources.\\
Relevant data on the BL Lac sample are reported in Tables \ref{tab:table1} and \ref{tab:table2}.\\
In Fig. \ref{fig:fig4} we plot the observed spectral indices against the redshifts for BL Lacs, as they have been derived from our spectral fits for all the sources (colours differentiate the HBLs from the IBLs).
The best-fit line was estimated with a chi squared test, obtaining a high value of $\chi^2_\nu=2.61$ due to the large scatter of the data.
The plot shows a clear correlation, that we interpret as the effect of the EBL photon-photon opacity at the high photon energies, as discussed in Sect. \ref{EBL} below.

\subsubsection{The BL Lac source PG 1553+113}

The BL Lac object PG 1553+113 had a significant impact in the analysis of \cite{2020MNRAS.493.1553G}, by contributing to the correlation signal of spectral indices against redshift that they found. This was particularly because of their choice to consider, among the published VHE spectra of the source, the one measured by HESS during a flare that showed a peculiarly hard spectral index \citep[][]{2008AA...477..481A}.

A first aspect of concern for this source is the lack of a robust spectroscopic measurement of the redshift. Of the various analyses mentioned in the TeVCat catalogue (all reporting values around $z=0.5$), we have found that by \cite{2022MNRAS.509.4330D} particularly accurate, $z=0.433$, and we will use this in the following. This is also consistent with the estimate by \cite{2007A&A...473L..17T}.

The source, observed during the 9-years of the Fermi satellite mission from 2008 to 2017, has shown a clear periodicity in luminosity with about constant period of $\sim 2$ years. \cite{2018ApJ...854...11T} interpret it in terms of a rare case of a binary system of super-massive black-holes of $10^8$ and $10^7\ M_\odot$, in which the smaller object periodically perturbs the jetted emission of the most massive one, producing particle acceleration via magnetic reconnection or other magneto-hydro-dynamical effects. \\
Of the four spectra mentioned in Table \ref{tab:table2}, that we individually analyzed, we have found the HESS 2005-2006 one reported by \cite{2008AA...477..481A} to show a bumpy irregular shape not fittable by a power-law, and because of that we have excluded it from our later analysis.


\subsection{The Flat-Spectrum Radio Quasar sample}

BL Lac and FSRQ sources belong to the same blazar category, as their multi-band Spectral Energy Distributions (SED) are very similar,  well described by double-peaked power-law functions. 

With respect to BL Lacs, FSRQs are intrinsically more luminous in all bands, and their spectrum is shifted towards lower frequencies. 
Also, emissions by the accretion disk and the broad-line clouds are evident in their spectra, with the consequence that some sources may have enhanced internal absorption by internal photon-photon interaction and $e^+e^-$ pair production with photons coming from the bright disk and the broad line region scattered towards the jet.
This mechanism, known as External Compton, would therefore be added to the Synchrotron Self Compton. 

For these reasons, while FSRQs are easily traceable at HE energies and have secure redshifts, they are more difficult to detect at VHE above $E_{0}=100$ GeV. Indeed, the FSRQs observed so far by IACTs are still limited to 9 sources. 
However, the motivation to include FSRQs in our analysis comes from their higher $z$ and secure redshift measurements.

The FSRQ sources analyzed in this paper are only 5, for which 10 good associated spectra are available (see data in Table \ref{tab:table3}). However, unlike the BL Lacs which are observed up to $z \sim 0.5$ at most, for the FSRQs there are 2 sources at $z> 0.9$. For such objects we expect very significant absorption effects by the EBL and likewise good tests of the ALP-photon mixing.


\subsection{Variability}

Blazars also show events of high variability approximately simultaneous in both synchrotron and Inverse Compton peaks. 
    
Most VHE blazars only show a factor of 2-3 in VHE flux variations, with notable episodes of rapid minute-scale events. 
Large-scale (factors of $>10$) flux variations are instead very rare. It is worth noting that the observed time scales for these smaller variations (days to years) often depend on the brightness of the objects in the VHE band, with shorter-duration variations only seen during isolated flaring episodes or for only the brightest objects. 
About a third of VHE AGN are detected during flaring events, easing their detection with the current moderate-sensitivity instrumentation.

\section{EBL photon-photon absorption corrections}
\label{EBL}

The first, zero-th order correction that we need applying to our collected VHE spectral data concerns the opacity of the Universe caused by the large volume density of low-energy photons and their interaction with those emitted by distant blazars.
This photon-photon interaction -- whose maximal cross-section occurs when the product of photon energies is equal to that of the electron and positron rest-energies -- and the relative gamma-ray destruction, are inevitable consequences of quantum mechanics \citep[][]{heitler1954}. 

To calculate such VHE opacity, the EBL spectral intensity contributed by cosmic sources and its time evolution have been modelled by various authors \citep[][among various others]{1998ApJ...494L.159S,2002AA...386....1K,2008A&A...487..837F,2010ApJ...712..238F,2011MNRAS.410.2556D,2012MNRAS.422.3189G,2016ApJ...827....6S,2018A&A...614C...1F}.

We adopt here for our corrections the EBL photon density model from the far-UV to the sub-millimeter ($\rm 0.1\ \mu m<\lambda < 1000\ \mu m$)  by \citet{franceschini2017,2018A&A...614C...1F}.
Their approach was to adopt a {\it backward evolution} model, starting from a detailed knowledge of the local luminosity functions of galaxies and active nuclei all-over the wavelength interval, and the tight constraints on how such functions evolve back in cosmic time made available by the large variety of deep multi-wavelength surveys. 
Once the redshift-dependent luminosity functions are determined 
based on all these data, an integral in luminosity gives the source emissivity and the EBL photon density as a function of wavelength and redshift.

Our adopted empirical approach to model the EBL absorption corrections offers maximal adherence to the observational data and is clearly to be preferred to alternative approaches relying on theoretical prescriptions about birth and evolution of galaxies and AGNs \citep[e.g.][]{2012MNRAS.422.3189G}.
Note also that, starting from 2008, the results by the most referred to EBL models have nicely converged, at least below $z\sim 1$ that is of interest for us here. With the implication that differences of only of the order of 10\% at most would be found by varying the adopted EBL model from one to the other, that is below the statistical uncertainties inherent in the analysis.

A second integral in redshift of the photon number density, calculated from $z=0$ to that of gamma-ray source $z_{\rm source}$ and properly weighted by the gamma-gamma cross-section, gives us the optical depth $\tau(\epsilon, z_{\rm source})$ as a function of the gamma-ray energy $\epsilon$. This is  related to the probability that a such VHE photon could have been absorbed during the path along the line-of-sight, a quantity that can be precisely calculated for both local and high redshift gamma-ray emitters.
All the details of the calculation can be found in \citet{2008A&A...487..837F} and \citet{franceschini2017}, while updated tables of the photon density and optical depths $\tau_{EBL}(\epsilon, z_{\text{source}})$ are reported in \citet{2018A&A...614C...1F}.

The intrinsic spectrum is then calculated as the observed one times the factor  $e^{\tau_{EBL}(\epsilon, z_{\text{source}})}$:
\begin{equation}
S_{\rm em}(\epsilon,z)=S_{\rm obs}(\epsilon,z)\times  e^{\tau_{EBL}(\epsilon, z_{\text{source}})},
\label{true1}
\end{equation}

Examples of absorption correction factors are illustrated in Fig. \ref{abs} for three characteristic values of the source redshift.
Fig. \ref{pg1553} also illustrates a conventional spectral correction for pure EBL-induced opacity on a VHE spectrum by VERITAS of the source PG 1553+113, together with ALP corrections to be discussed later.

\subsection{Photon-photon correction uncertainties}
\label{EBLunc}

Uncertainties in the corrections for EBL photon-photon absorption have to be evaluated in order to properly assess the significance of our results.
\cite{2008A&A...487..837F} and later publications by the team did not address them, with the understanding that they are so small not to impact significantly in the analyses of VHE spectra because of their large statistical errors.

A different approach has been followed by \cite{2016ApJ...827....6S}, that estimated global uncertainties in their EBL intensity and optical depth estimates as large as a magnitude (factor $\times 2.5$) on average. Their analysis is based on published luminosity functions for galaxies selected at various wavelengths, integrated to calculate the background intensity, and then fitted with simple analytic functions. This allowed them to perform an error propagation analysis to get their final uncertainties, but these are so large due to the typically large statistical errorbars in the luminosity functions (from small samples in small areas). 

Our independent model has exploited a much more extensive multi-wavelength dataset, particularly including the deepest galaxy number counts at all wavelengths. The latter have very small errors because involving large samples of sources on various extended sky areas, to very faint limits.  
Remember that the source background intensity at low-redshifts $I_\nu$ is essentially determined by a direct integral of the differential counts $N(S_\nu)$, 
$$ I_\nu\propto \int S_\nu \times N(S_\nu)dS_\nu
$$
and the photon-photon optical depth is just proportional to this intensity.
Because our model, in addition to the number counts,
involves the source redshift distributions and luminosity functions, also the redshift dependence of the background flux is accounted for.
As discussed in \cite{2000MNRAS.312L...9M}, and later confirmed by many other published results, the uncertainties in the individual wavelength bins are of the order of $\delta I_\nu/I_\nu\sim 0.2$, thanks to the fact that counts at all wavelengths already converge at faint fluxes and leave no significant room to undetected sources.
Consequently, the maximum uncertainty after integration in wavelength to get the total intensity and, proportionally, the opacity $\tau_{EBL}(\epsilon, z_{\text{source}})$ is less than that.

In our following analysis we adopt two values for the relative uncertainty in the photon-photon opacity:
\begin{equation}
\cfrac{\delta \tau_{EBL}}{\tau_{EBL}}=0.1,\ \  0.2 
\label{tauunc}
\end{equation}
to represent a best-guess and a conservative estimate, respectively. The corresponding error in the optical depth $\sigma_{EBL} = \cfrac{\delta \tau_{EBL}}{\tau_{EBL}}\tau_{EBL}$ 
has to be quadratically added to the statistical errorbars of the spectral energy bins.
From the error propagation formula 
$$
\sigma(S_{em})=\sqrt{\left(\frac{d S_{em}}{d S_{obs}} \sigma_{stat}\right)^2+\left(\frac{d S_{em}}{d\tau_{EBL}} \sigma_{EBL}\right)^2},
$$
where $\sigma_{stat}$ is the statistical error on the observed fluxes, we get
\begin{equation}
\sigma(S_{em})=\sqrt{(e^{\tau_{EBL}} \sigma_{stat})^2 + (S_{obs} e^{\tau_{EBL}}\ \sigma_{EBL})^2}.
\end{equation}
This is the total uncertainty to be associated to the source emitted flux $S_{em}$ in eq. \ref{true1}.
Results for both assumed values of $\sigma_{EBL}$ will be reported below.

\begin{figure*}
\includegraphics[width=0.75\textwidth,height=0.45\textwidth]{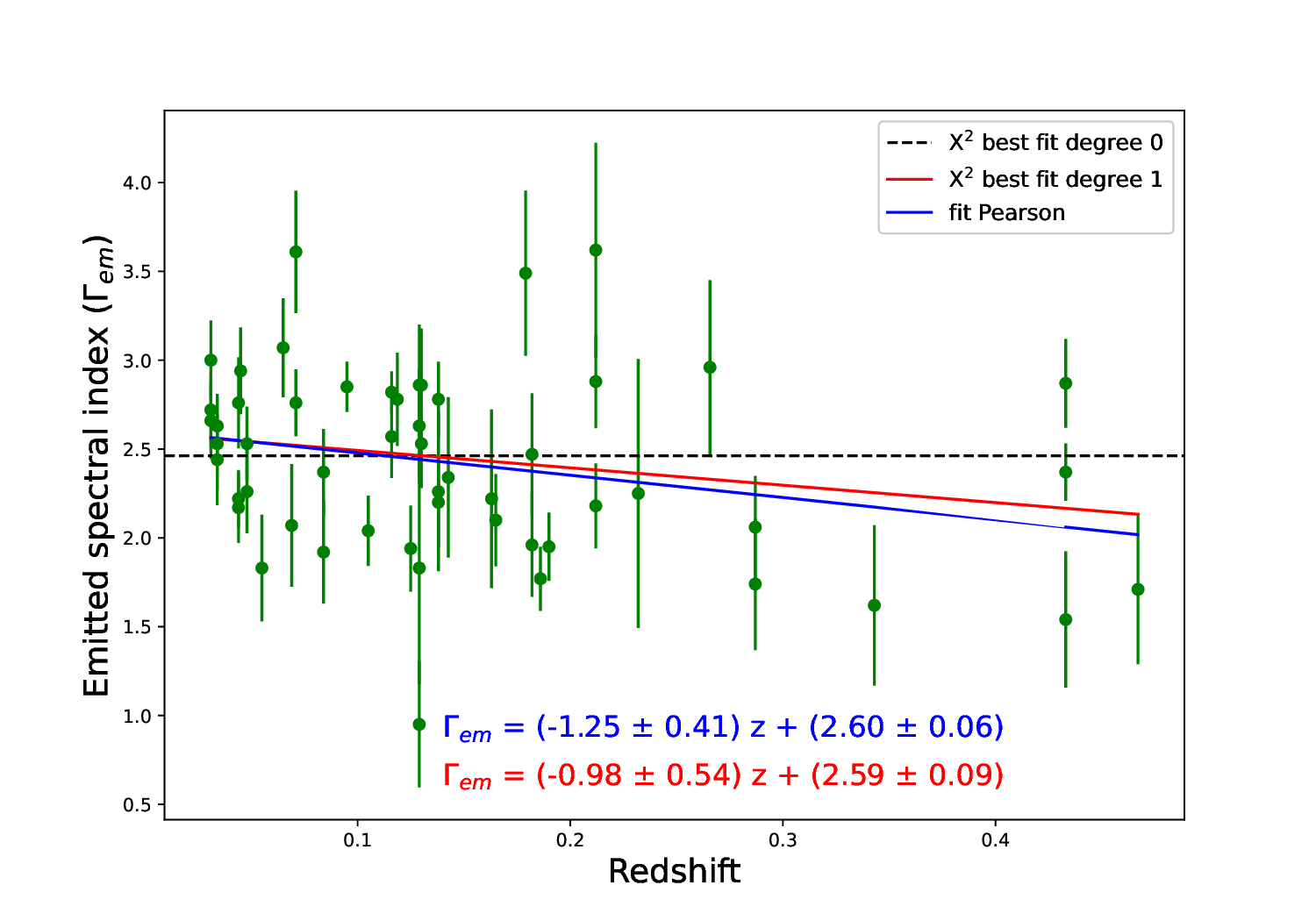}
\caption{Emitted spectral indices of HBL BL Lac sources after correction for EBL-only opacity, plotted against the source redshifts. A dependence on $z$ is apparent, whose significance is quantified in Table \ref{tab:table4}.
}
\label{fig:true}
\end{figure*}

\begin{figure*}
\includegraphics[width=0.75\textwidth,height=0.45\textwidth]{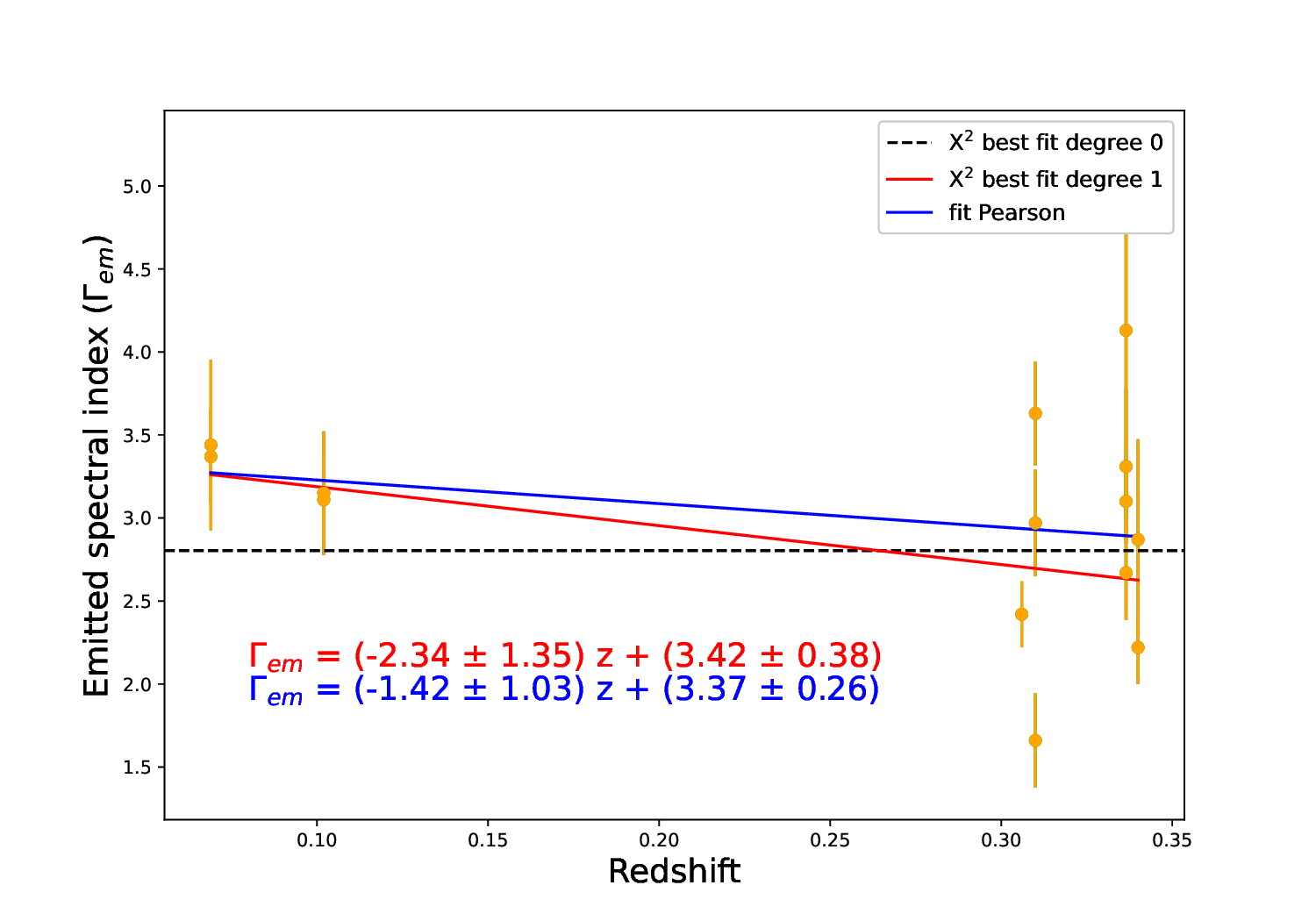}
\caption{Emitted spectral indices of IBL BL Lac sources after correction for EBL-only opacity against redshift. See also Fig. \ref{fig:true} and Table \ref{tab:tableIBL}. 
If we do not consider the two sources cataloged "blazar" and "BL Lac" in TeVCat, the two linear regression lines would be even more steeper.
}
\label{fig:trueIBL}
\end{figure*}

\begin{figure}
\includegraphics[width=0.5\textwidth,height=0.40\textwidth]{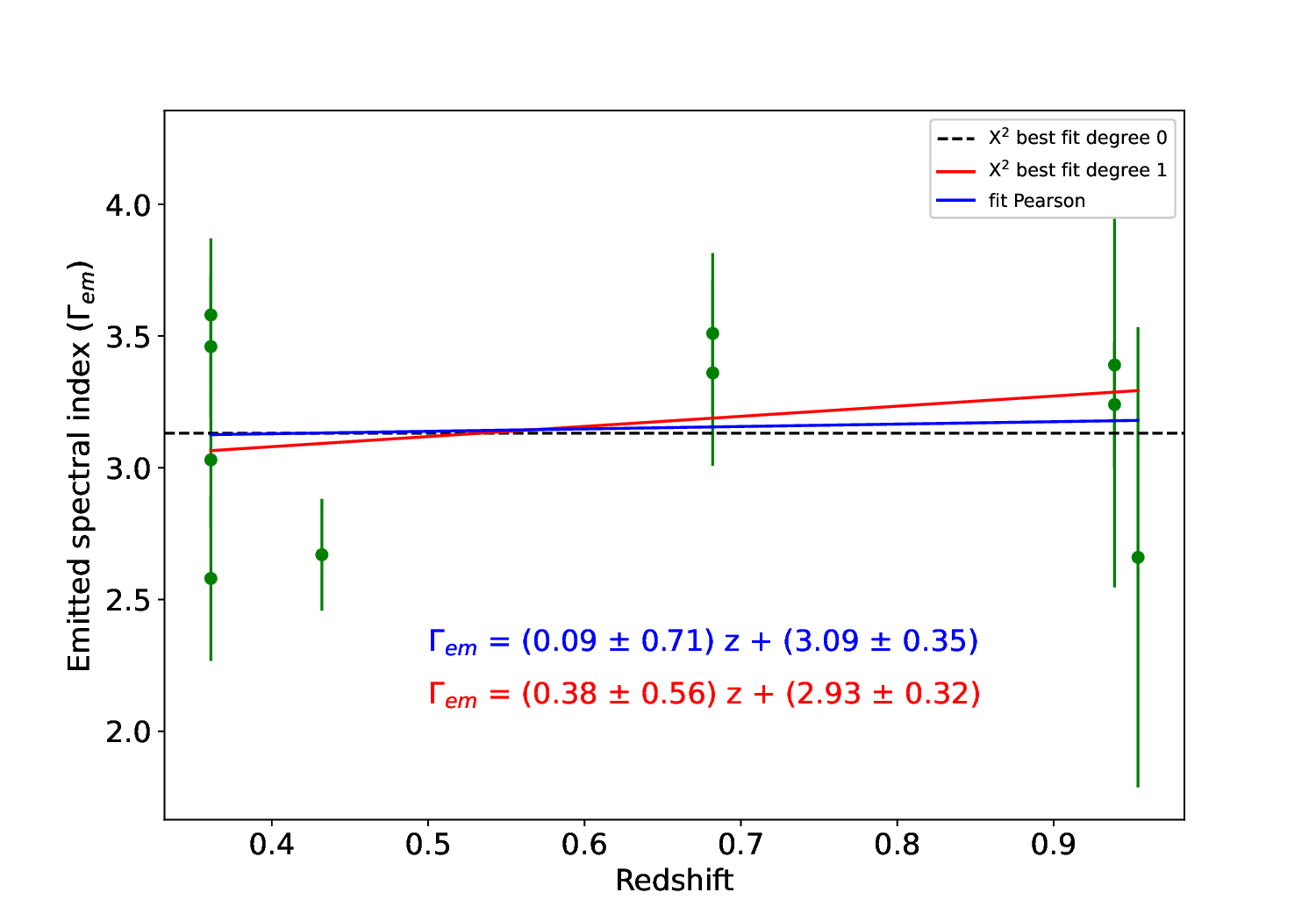}
\caption{Same as Fig. \ref{fig:true}, for FSRQs. Results of the tests in Table \ref{tab:table5}.}
\label{fig:FSRQ}
\end{figure}

\section{Statistical analysis: EBL correction}
\label{EBLcorrection}

Having calculated the spectral EBL-based corrections for all sources at their redshifts $z_{\text{source}}$, all spectral data-points are corrected to obtain the intrinsic spectrum in the source-frame as in eq. \ref{true1}, 
and then newly fitted with power-law functions 
\begin{equation}
S_{\rm em}(\epsilon,z) \propto \epsilon^{-\Gamma_{\rm em}}
\label{true2}
\end{equation}
to obtain the source intrinsic spectral indices $\Gamma_{\rm em}$, together with the irrelevant normalization constants. These spectral indices are plotted as a function of the redshift in Figs. \ref{fig:true} and \ref{fig:trueIBL}, separately for the HBL and IBL objects.
We see that both plots reveal some residual anti-correlations with redshift, meaning a tendency of the EBL factors to over-correct the spectra at the high energies.
The results of a similar analysis carried out for the few FSRQs are reported in Fig. \ref{fig:FSRQ}.

The entity and significance of these correlations have been tested in various ways, as detailed below.

\begin{table*}
\begin{center}
\caption{Tests of the correlations of $\Gamma_{\rm em}$ spectral indices with $z$ for HBL BL Lacs.}\label{tab:table4}
\begin{tabular}{lccccc} 
\hline
Test           & Pearson     & Kendall                  & Spearman   &  Horizontal fit   &   Linear fit        \\
& & &       &  $\chi^2_{\nu,0}$ (ndof=53)       &      $\chi^2_{\nu,1}$ (ndof=52)         \\
\hline                                                  
Observed   &  $0.57\pm0.05 $     & $0.37\pm 0.04$  & $0.54\pm 0.06$  &  4.41     & 2.55 \\
P-value  &     $0.005\% $     &     $0.03\% $     &       $0.02\%$  & &  \\
\hline                                                  
EBL-corrected $\left[\sigma(\tau_{EBL})=0.1\tau_{EBL}\right]$ &  $-0.24 \pm 0.07$     & $-0.17 \pm 0.05$  & $-0.25 \pm 0.07$   &  3.14     & 3.01  \\
P-value   &     $12\% $     &     $10\% $     &       $10\% $  & &  \\
\hline                                                  
EBL-corrected $\left[\sigma(\tau_{EBL})=0.2\tau_{EBL}\right]$ &  $-0.23 \pm 0.07$     & $-0.17 \pm 0.05$  & $-0.23 \pm 0.07$   &  3.07     & 2.94  \\
P-value   &     $14\% $     &     $12\% $     &       $12\% $  & &  \\
\hline                                                  
ALP-corrected ($\xi=1.0$)   &  $0.00\pm0.07$ & $-0.05\pm0.05$ &    $-0.07\pm0.08$  &   2.82     & 2.87\\
P-value   &     $70\% $     &     $56\% $     &       $57\%$   & & \\
\hline                                                  
ALP-corrected  ($\xi=2.0$)  &   $0.09\pm0.07$     & $0.02\pm0.05$  & $0.04\pm0.08$   &   2.69     & 2.69 \\
P-value   &     $53\% $     &     $65\% $     &       $63\%$   & & \\
\hline                                                  
\hline
\end{tabular}
\end{center}
\raggedright
\footnotesize 
\texttt{Col.1}: Data sample used for the statistical tests; \texttt{Col.2-3-4}: Correlation coefficients and probability values obtained from different tests using Monte Carlo simulations; \texttt{Col.5}: Reduced chi-squared obtained by horizontal fit; \texttt{Col.6}: Reduced chi-squared obtained with the linear fits shown in Fig. \ref{fig:fig4}, \ref{fig:true},\ref{fig:ALP1} and \ref{fig:ALP2}.
For the purely EBL-corrected values we report in the second and third row results assuming our two reference uncertainties $\sigma(\tau_{EBL})/\tau_{EBL}=0.1$ and 0.2 for the EBL corrections, as indicated, while
for the ALP-corrections only the values for $\sigma(\tau_{EBL})=0.2\tau_{EBL}$.

\end{table*}

\begin{table*}
\begin{center}
\caption{Tests of correlation for IBL BL Lacs.}\label{tab:tableIBL}
\begin{tabular}{lccccc} 
\hline
Test           & Pearson     & Kendall                  & Spearman   &  Horizontal fit   &   Linear fit        \\
& & &       &  $\chi^2_{\nu,0}$ (ndof=13)       &      $\chi^2_{\nu,1}$ (ndof=12)         \\
\hline                                                  
Uncorrected   &  $0.15\pm0.19 $     & $0.08\pm 0.16$  & $0.11\pm 0.21$  &  1.57     & 1.63 \\
P-value  &     $53\% $     &     $55\% $     &       $55\%$  & &  \\
\hline                                                  
EBL-corrected $\left[\sigma(\tau_{EBL})=0.2\tau_{EBL}\right]$ &  $-0.23 \pm 0.17$     & $-0.18 \pm 0.14$  & $-0.23 \pm 0.18$   &  3.65     & 3.15  \\
P-value   &     $46\% $     &     $42\% $     &       $44\% $ & &  \\
\hline                                                  
ALP-corrected ($\xi=1.0$)   &  $-0.18\pm0.18$ & $-0.15\pm0.14$ &    $-0.18\pm0.19$  &   2.36     & 2.19\\
P-value   &     $53\% $     &     $48\% $     &       $50\%$ & &   \\
\hline                                                  
ALP-corrected  ($\xi=2.0$)  &   $-0.13\pm0.18$     & $-0.12\pm0.14$  & $-0.15\pm0.19$   &   1.96     & 1.93 \\
P-value   &     $59\% $     &     $51\% $     &       $52\%$  & &  \\
\hline                                                  
\hline
\end{tabular}
\end{center}
\raggedright
\footnotesize 
\texttt{Col.1}: Data sample used for the statistical tests; \texttt{Col.2-3-4}: Correlation coefficients and probability values obtained from different tests using Monte Carlo simulations; \texttt{Col.5}: Reduced chi-squared obtained by horizontal fit; \texttt{Col.6}: Reduced chi-squared obtained with the linear fits shown in Fig. \ref{fig:fig4} and \ref{fig:trueIBL}. 
 For the purely EBL-corrected values and for the ALP-corrections we used only the uncertainties $\sigma(\tau_{EBL})=0.2\tau_{EBL}$.

\end{table*}

\begin{table*}
\begin{center}
\caption{Tests of correlation for FSRQs.}  \label{tab:table5}
\begin{tabular}{lccccc} 
\hline
Test           & Pearson     & Kendall                  & Spearman   &  Horizontal fit   &   Linear fit        \\
& & &       &  $\chi^2_{\nu,0}$ (ndof=9)       &      $\chi^2_{\nu,1}$ (ndof=8)         \\

\hline                                                  
Uncorrected   &  $0.39 \pm 0.22$     & $0.22 \pm 0.18$  & $0.26 \pm 0.23$ & 2.06 & 1.97 \\
P-value   &    30\%         &          43\%               &          47\%              &           &              \\
\hline                                                  
EBL-corrected $\left[\sigma(\tau_{EBL})=0.2\tau_{EBL}\right]$ &  $0.05 \pm 0.31$     & $0.00 \pm 0.22$  & $0.00 \pm 0.28$ & $1.61$  & 1.70\\
P-value   &     52\%              &  55\%                        &      57\%                  & & \\
\hline                                                  
ALP-corrected ($\xi=1.0$) &   $0.25 \pm 0.29$     & $0.15 \pm 0.22$  & $0.18 \pm 0.27$ & $1.85$ & 1.66 \\
P-value       &    43\%             &    49\%                      &          53\%   & &                       \\
\hline                                                  
ALP-corrected ($\xi=2.0$) &   $0.19 \pm 0.30$     & $0.10 \pm 0.22$  & $0.11 \pm 0.27$ & 1.75 & 1.73  \\
P-value       &    47\%  & 53\%                  &  57\%    &           &                         \\
\hline                                                  
\hline
\end{tabular}
\end{center}
\raggedright
\footnotesize 
\texttt{Col.1}: Data sample used for the statistical tests; \texttt{Col.2-3-4}: Correlation coefficients and probability values obtained from different tests using Monte Carlo simulations; \texttt{Col.5}: Reduced chi-squared obtained by horizontal fit; \texttt{Col.6}: Reduced chi-squared obtained with a linear fit shown in Fig. \ref{fig:FSRQ} and \ref{fig:ALP3}.
For the purely EBL-corrected values and for the ALP-corrections we used only the uncertainties $\sigma(\tau_{EBL})=0.2\tau_{EBL}$.
\end{table*}

\begin{itemize}

    \item 
\textbf{The Pearson correlation test.} 
\\The Pearson's correlation coefficient is a measure of linear correlation between two sets of data. It is the ratio between the covariance of two variables and the product of their standard deviations. Thus it is essentially a normalized measurement of the covariance, with values ranging between $-1$ and $+1$ for perfect negative and perfect positive linear correlation, respectively. As for the covariance itself, the measure can only reflect linear dependencies of variables, and ignores other types of relationship. 

The test offers a probability value (p-value) roughly indicating the chance of an uncorrelated data-set with the same numerology producing a Pearson correlation value at least as extreme as that of the sample under analysis. More precisely, for a given sample with correlation coefficient $r$, this p-value is the probability that $ |r'|$ for a random sample drawn from the population with zero correlation would be greater than or equal to $|r|$.
    \item 
\textbf{The Spearman correlation test.}
 \\The Spearman's correlation coefficient uses the same formula as the Pearson's test, simply applying it to the ranks of the variables. Unlike the Pearson correlation, the Spearman correlation does not assume that both data-sets are normally distributed. Furthermore, the Spearman's test assesses monotonic relationships, whether linear or not, so it is more general than Pearson's.
 The test provides a correlation significance (p-value) not accounting for the measurement uncertainties in the data.

    \item 
\textbf{The Kendall correlation test.} 
\\The Kendall's correlation test uses a totally different formula than Pearson and Spearman, based on the number of \textit{concordant} and \textit{discordant} pairs of data between two data-sets: two distinct pairs $(x_i, y_i)$ and $(x_j, y_j)$ are said:\\
- \textit{concordant} if ($x_i > x_j$ and $y_i > y_j$) or ($x_i < x_j$ and $y_i < y_j$)\\
- \textit{discordant} if ($x_i > x_j$ and $y_i < y_j$) or ($x_i < x_j$ and $y_i > y_j$).\\
The formula used for this test is:
\begin{equation}
\tau=\frac{C-D}{C(n,2)}
\label{true3}
\end{equation}
where C is the number of concordant and D the number of discordant pairs. $C(n,2)$ is the number of the possible ways of selecting distinct pairs $(x_i, y_i)$ and $(x_j, y_j)$.
As for the Spearman's correlation test it is sufficient that the relationship between the variables is monotonic.
 Also this test provides a correlation significance (p-value) not accounting for the measurement uncertainties in the data.\\

    \item 
\textbf{$\chi^2$ and F-tests}
\\
We used the Chi-Square as a goodness-of-fit to test the various relations between the $\Gamma_{\rm em}$ spectral indices and the redshifts, under various conditions.
The algorithm used is the Levenberg-Marquardt Algorithm (LMA), a generally efficient and reliable numerical procedure for finding local minima in the chi-square $\chi^2$ values. The application requires to provide initial values to the routine, from which it can start exploring the space of $\chi^2$ values. 

In addition to linear coefficients and related errors, the program returns two factors: the normalized covariance $c$ and the reduced chi-square ($\chi^2_\nu$) (ratio between the $\chi^2$ and the number \textit{ndof} of degrees of freedom of the system). 
The reduced chi-square is used to check if the experimental data are well expressed by a theoretical model described by the regression line function.

Finally, we partly used the F-test to evaluate the significance of adding a parameter to our model, like would be a linear correlation with $z$ in the data against a constant spectral-index distribution. The test is based on the ratios of the reduced chi-square values $\chi^2_\nu$ for the two models.
\end{itemize}

\subsection{Estimating the global uncertainties}
\label{uncertain}

To account for the uncertainties in our measured spectral indices we used a Monte Carlo approach. To this end, for each individual source's spectral index $\Gamma_{\rm em}$, we have randomly modified it assuming an underlying Gaussian distribution of values with standard deviation equal to the error in the index measurement (note that included here are all the statistical and systematic errors of the source's binned flux measurements, and those related to the EBL corrections), so obtaining many randomly-generated samples (no error is assumed for the redshift). We then calculated the Pearson/Spearman/Kendall correlation coefficients for each simulated data-set, generating a distribution of the correlation coefficients from which an improved estimate of the correlation significance (P-value) is calculated. This, properly accounting for the $\Gamma_{\rm em}$ measurement's total uncertainties, will be our considered P-value in the following.
In the same way, we calculated the errors in the regression lines from each data-set using the Pearson formulae, as reported in the figures. 

All the results of our statistical analysis are reported in Tables \ref{tab:table4}, \ref{tab:tableIBL}, and \ref{tab:table5} for the HBL, IBL, and the FSRQ samples, respectively. The first row in all tables refer to the observed $\Gamma_{\rm obs}$:
as expected, the observed data for BL Lacs display highly significant positive correlation, an evidence that the spectral index $\Gamma_{\rm obs}$ increases with redshift, as an effect of an increasing photon-photon opacity with distance.

The second and third rows report the EBL-corrected $\Gamma_{\rm em}$ values assuming our two reference uncertainties $\sigma(\tau_{EBL})/\tau_{EBL}=0.1$ and 0.2 for the EBL corrections
(for the ALP-corrections in rows forth and fifth only the values for the more conservative $\sigma(\tau_{EBL})=0.2\tau_{EBL}$). 

The EBL spectral corrections, in their turn, appear to produce a slight
anti-correlation of $\Gamma_{\rm em}$ with redshift for both HBLs and IBLs, that has to be interpreted.
The P-value probabilities that the data come from an un-correlated distribution show a lack of significance (large P-values) for IBLs and FSRQs, while for the more numerous HBL sample they get to the level ($\simeq 12\%$ to 14\%) corresponding to $\sim 1.6\sigma$ significance.

The $\chi^2$ test turns out to be of limited use, because the spectral-index data have a large dispersion. Even the best-fit linear regressions in columns 6 have large reduced chi-squares. For basically this reason, also the F-tests are essentially of no use to single out the best models: for example, the F-test favours a linear increase of  $\Gamma_{\rm obs}$ with redshift against a constant distribution only at the $\sim 1\%$ level, while the other tests reveal evidence for correlation in a much more significant way.

\begin{figure*}
\includegraphics[width=0.75\textwidth,height=0.5\textwidth]{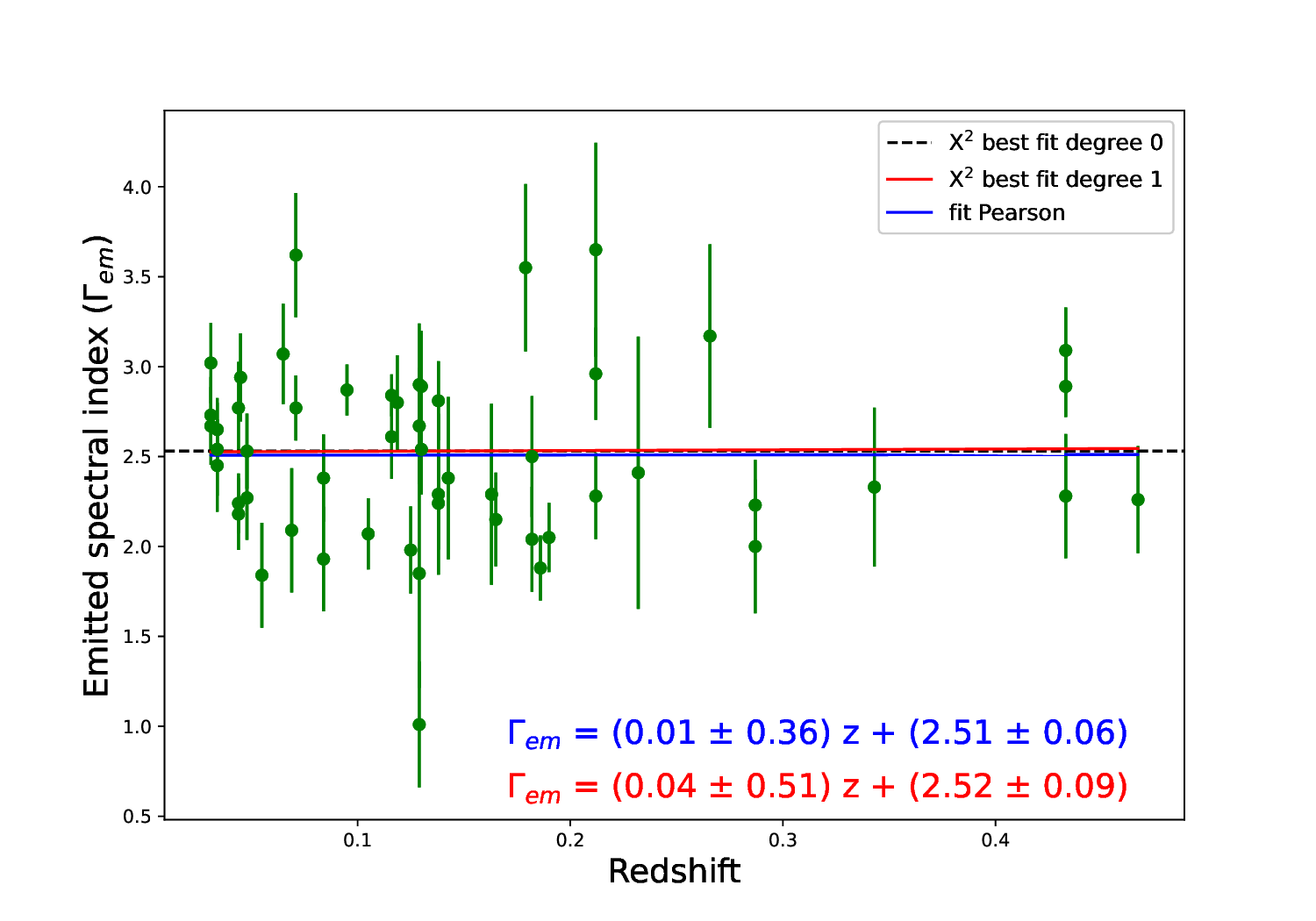}
\caption{ Plot of the emitted spectral indices of HBL BL Lac sources with ALP-corrected opacity against the source redshifts. The dimensionless ALP parameter is set here to $\xi=1$. 
The small residual dependence on redshift is quantified in Table \ref{tab:table4}.
}
\label{fig:ALP1}
\end{figure*}

\begin{figure*}
\includegraphics[width=0.75\textwidth,height=0.5\textwidth]{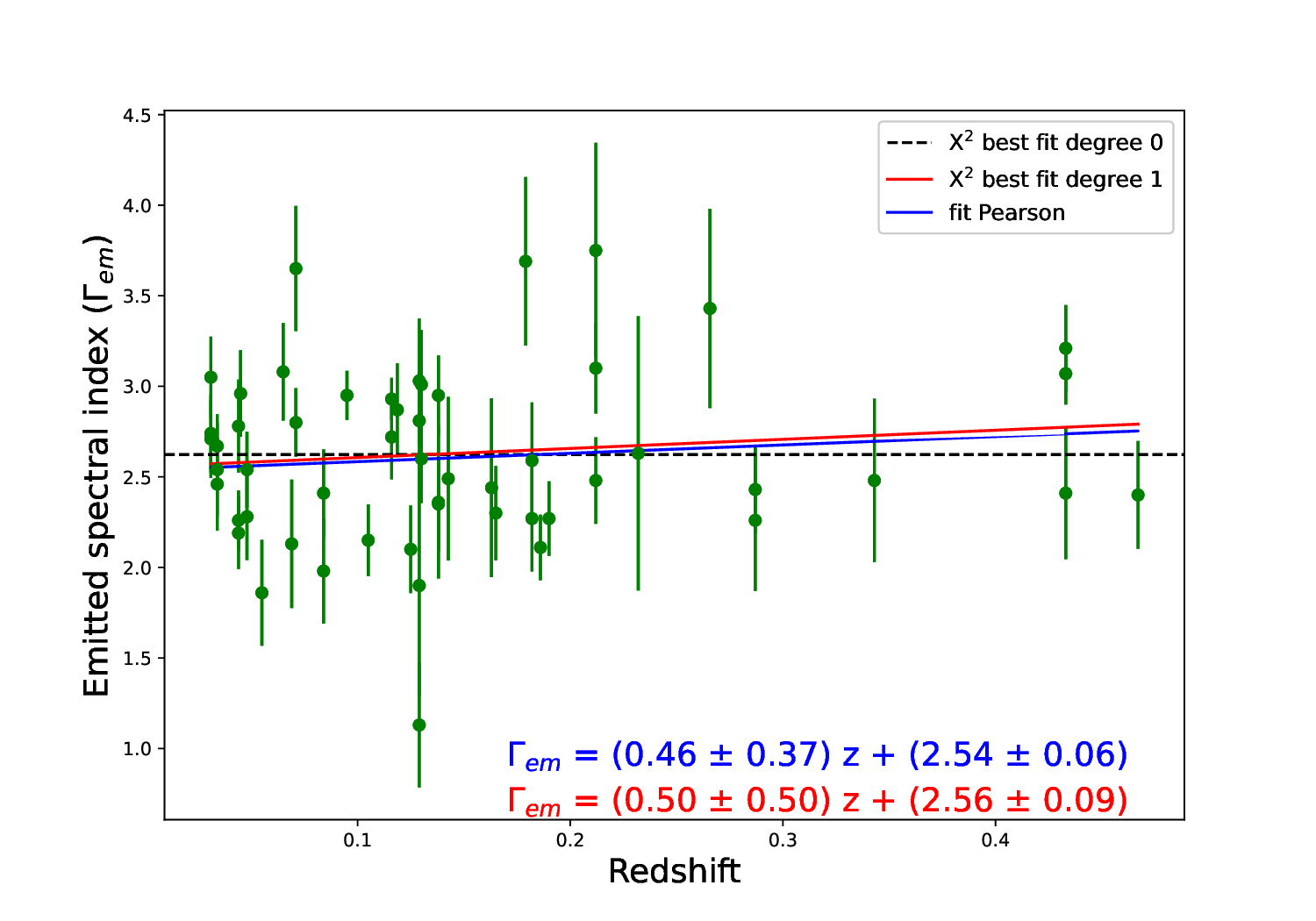}
\caption{ Same as in Fig. \ref{fig:ALP1}, with the ALP parameter set to $\xi=2$. 
}
\label{fig:ALP2}
\end{figure*}

In conclusion, the residual correlations of $\Gamma_{\rm em}$ with redshift, although not so significant due to our limited statistics, may have various explanations.

One is that our samples might suffer a bias due to the limited sensitivity of the current IACT telescopes, such that only sources with the hardest VHE spectra would be bright enough to be well observable above $100 \ \rm GeV$. This possibility has been discussed and excluded by \cite{2020MNRAS.493.1553G}, who found no correlation of the spectral indices with the source luminosity.

Another possibility is that our adopted EBL model over-corrects for the photon-photon opacity, due to an excess of low-energy photons in the optical-UV: these are the photons mostly responsible for the opacity correction, as it is evident from consideration of the VHE energy ranges in Figs. \ref{fig:fig2} and \ref{fig:fig3} and of the approximate relation where maximum opacity occurs:
\begin{equation}
    \frac{ \epsilon}{\rm TeV}\simeq \frac{\lambda}{\rm \mu m} .
\end{equation}
But no evidence can be found in the literature that our model systematically exceeds the EBL intensity between 0.1 and 1 $\rm \mu m$, see e.g. \cite{2011MNRAS.410.2556D} and \cite{2016ApJ...827....6S}.

Finally, an interesting possibility to consider is that we are missing so far something fundamental in the physical process of photon interaction, warranting further inspection in terms of e.g. a possible photon-to-ALP mixing, overall reducing the cosmic VHE opacity.
This will be addressed in the next section.


\section{Statistical Analysis: ALP Correction}
\label{corrALP}

\subsection{The physical process}
\label{physical}

 The ALP theory is characterized by a Lagrangian term:
\begin{equation}
{\cal L}_{a\gamma}=g_{a\gamma\gamma} \, {\bf E}\cdot {\bf B} \, a
\label{lagrangian}
\end{equation}
where ${\bf E}$ and ${\bf B}$ are the electric and magnetic components of the electromagnetic field, $g_{a\gamma\gamma}$ is the two-photon coupling constant and $a$ is the ALP field. 
This implies that ALPs and photons can oscillate into each other in the presence of an external electromagnetic field, in a process know as photon-ALP mixing. 
 
The key-point in the ALP scenario is that ALPs do not significantly interact with the EBL photons, in spite of the fact that they couple to two photons, because of the very low interaction cross-section (see e.g. \citealt[][]{grExt}).
Once generated by the source, VHE photons travel across the Universe by crossing regions characterized by intergalactic magnetic fields $\bf B$ with different orientations, as well as EBL photons and, depending on the orientation of $\bf B$, they are (or are not) transformed into ALPs. So, for a fraction of their travel to the Earth, in the ALP phase, they travel undisturbed, and may be back transformed into photons by interaction with local $\bf B$-fields.

In summary, including this ALP effect, the optical depth $\tau(\epsilon, z_{\text{source}})$ in Eq. \ref{true1} would be replaced by the effective optical depth $\tau^{\rm ALP}(\epsilon, z_{\text{source}})$ which is always smaller when EBL absorption is larger: the ratio $\tau^{\rm ALP}(\epsilon, z_{\text{source}})/\tau(\epsilon, z_{\text{source}})$ turns out to be a monotonically increasing function of both $\epsilon$ and $z$. Eq. \ref{true1} then becomes:
\begin{equation}
S_{\rm ALP}(\epsilon,z)=S_{\rm obs}(\epsilon,z)\times  e^{\tau^{\rm ALP}(\epsilon, z_{\text{source}})} .
\label{Salp}
\end{equation}

So, the main consequence of photon-ALP oscillations is to reduce the EBL absorption, hence increasing the cosmic transparency above $E_0\geq500$ GeV. We used in this work the same model of \cite{2020MNRAS.493.1553G}: a domain-like network, where the magnetic field $\bf B$ is homogeneous over a domain of size $L_{\rm dom}$ (the $\bf B$ coherence length). In the present paper some different input parameters are however used.  

The first important parameter in the ALP model is the intensity and topology of the cosmic magnetic fields $\bf B$. The photon to ALP transfer functions could be easily calculated if we knew the exact configuration of the magnetic field along the line of sight to a gamma-ray source. Unfortunately, our knowledge of it is very limited \citep[][]{2001PhR...348..163G}, and not even the approximate strength and coherence length are known. We have to rely on model predictions to assess the potential impact of photon-ALP oscillations on the photon propagation. According to \cite{1968Natur.217..326R} and  \cite{1969Natur.223..936H}, the existence of energetic quasar outflows should give rise to intergalactic magnetic fields, since the ejected material is ionized and magnetic flux lines are frozen in. In these models the outflows form bubbles with magnetic fields that are typically spread over $\sim$ 4 Mpc with field strengths of the order of 1 nG. Another possibility was put forward by  \cite{1999ApJ...511...56K}, which consists in galactic super-winds emitted by primeval galaxies ($z > 6$) magnetizing the extragalactic space and producing fields in the $0.1\ {\rm nG} < B < 1\ \rm nG$ range on the Mpc scale, in agreement with observations of Lyman-alpha forest clouds \citep[][]{1995AJ....109.1522C}, Faraday rotation measures \citep[][]{2016PhRvL.116s1302P} and other constraints \citep[][]{2013A&ARv..21...62D}.
An upper limit of 1.7 nG was also placed on the extragalactic magnetic field intensity by \cite{2016PhRvL.116s1302P}. For these reasons $B$ values around 1 nG were adopted and incorporated into our model parameters below.

\begin{figure*}
\includegraphics[width=0.75\textwidth,height=0.5\textwidth]{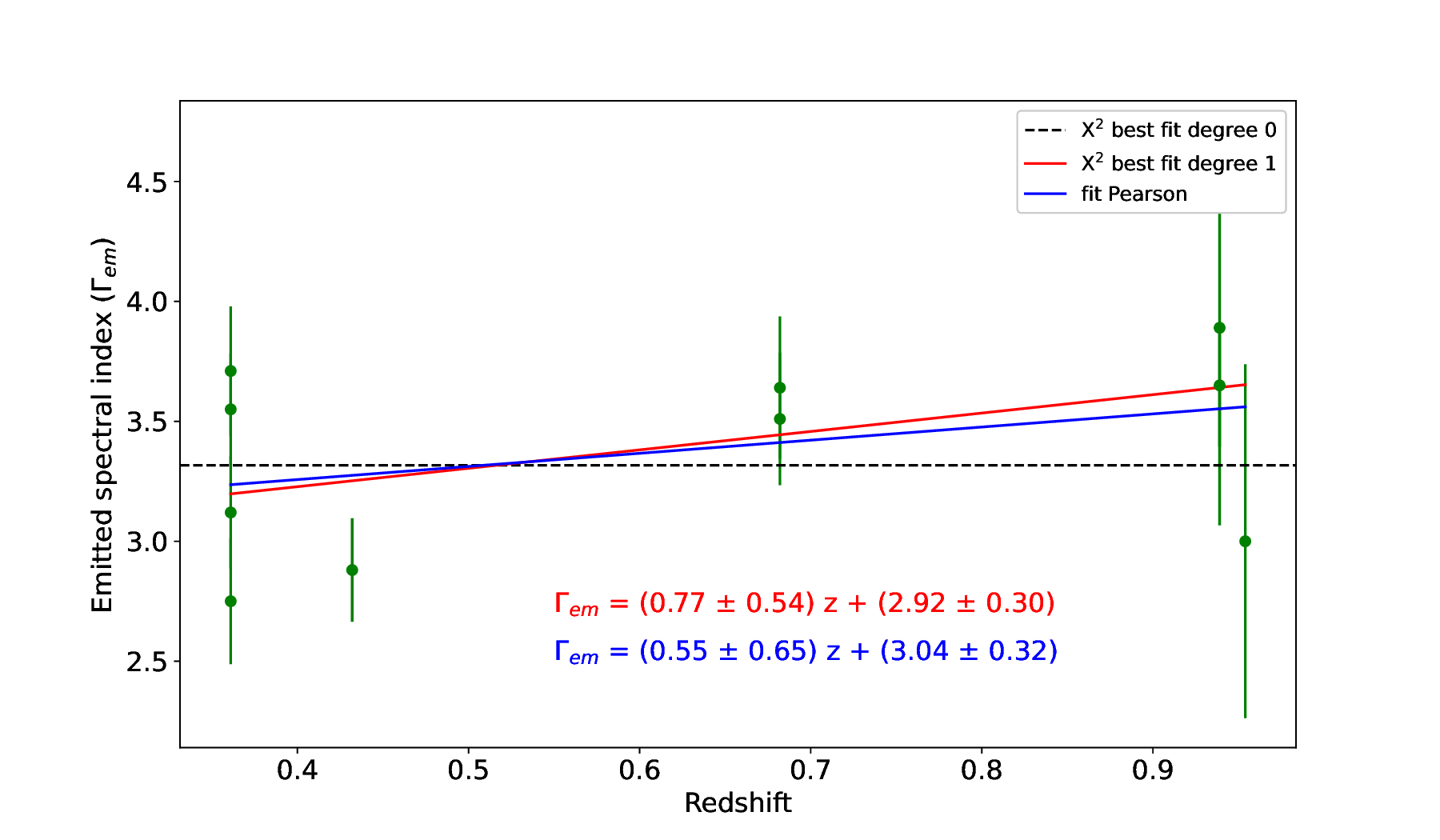}
\caption{ Plot of the emitted spectral indices of FSRQ sources with ALP-corrected opacity against the source redshifts. The dimensionless ALP parameter is set here to $\xi=1$. 
The small residual dependence on redshift is quantified in Table \ref{tab:table5}.
}
\label{fig:ALP3}
\end{figure*}


Our ALP correction prescription, still using the EBL model as discussed in Sect. \ref{EBL}, assumes that the $\bf B$ strength changes only slightly in all domains, with its direction changing randomly from one domain to the next. 
As previously mentioned, $L_{\rm dom}$ is the domain length along the line of sight. \cite{2020MNRAS.493.1553G} suggest values in the range $1 \, {\rm Mpc}\leq L_{\rm dom}\leq 10 \, \rm Mpc$. 
Working in the redshift space, we have assumed $\Delta z=0.001$, for which at the redshift $z$ we have 
$L_{\rm dom}(z,z+\Delta z)\simeq2960 \ln \left\{ [{1+1.45(z+\Delta z)}]/[{1+1.45 z}]\right\}$, corresponding to $L_{\rm dom}=4.29$ Mpc at $z=0$ and 1.9 Mpc at $z=0.9$.

The magnetic flux lines can be thought as frozen inside the intergalactic medium because of the high conductivity. Therefore, conservation of the $\bf B$ energy density during the cosmic expansion scales $\propto (1+z)^4$ and the intensity $B\propto (1+z)^2$.

The ALP mass $m_{\rm ALP}$, like the plasma frequency $\omega_{\rm pl}$, enter into the mixing matrix, quantifying the conversion probability of photon to ALP and vice versa (see e.g. \citealt{grSM} for details about the parameter's meaning and calculation).

Overall, our adopted physical parameter quantifying the relevance of the ALP effect is 
\begin{equation}
\xi\equiv \left(\frac{B}{1 \, {\rm nG}}\right) \left(g_{a\gamma\gamma} \cdot 10^{11}\ {\rm GeV}\right).
\label{xi}
\end{equation}
Note in any case that there is not a simple linear relation, but a rather complex one between $\xi$ and $\tau^{\rm ALP}$. The ALP correction increases from $\xi=0$ (corresponding to a pure EBL optical depth $\tau$) up to values of $\xi\simeq 1$ to 2, where the correction is maximal, corresponding to a maximal photon-ALP conversion probability in the single domain of the order of 10-50\%, with many conversions taking place along the entire journey of the  photon-ALP beam. For $\xi>2$ the ALP effect saturates. We defer to \cite{2011PhRvD..84j5030D} and \cite{2020MNRAS.493.1553G} for a more extensive discussion of the process.

In the present paper we have limited ourselves to the two representative values of $\xi=1.0$ and $\xi=2.0$.
Spectral fits denoted as ALP in Fig. \ref{pg1553} show examples based on ALP opacity corrections with $\xi=1.0$ and $2.0$, implying a photon-photon opacity lower than expected for the conventional EBL correction. The values of $g_{a\gamma\gamma}$ (inside the parameter $\xi$ of Eq. \ref{xi}) and $m_{\rm ALP}$ have been taken within currently acceptable ranges \citep{cast,straniero,fermi2016,payez2015,berg,conlonLim,meyer2020,limFabian,limJulia,limKripp,limRey2}, whose firmest bound is represented by $g_{a \gamma \gamma} < 0.66 \times 10^{- 10} \, {\rm GeV}^{- 1}$ for $m_{\rm ALP} < 0.02 \, {\rm eV}$ at the $2 \sigma$ level from no detection of ALPs from the Sun~\citep{cast}. In particular, $g_{a \gamma \gamma}$ has been set to $\mathcal{O}(10^{- 11}) \, \rm GeV^{-1}$ and $m_{\rm ALP}$ to $\mathcal{O}(10^{- 10}) \, \rm eV$ in the present study.

\subsection{Effects of a possible photon-ALP mixing}

The effects of opacity corrections, based on photon-ALP mixing, in the distribution of intrinsic VHE spectral indices $\Gamma_{\rm em}$ as a function of redshift are shown in the plots for the HBL BL Lacs in Figs. \ref{fig:ALP1} and \ref{fig:ALP2}.
The two figures refer to the two different values of the ALP parameter of Eq. \ref{xi}, $\xi=1$ and $\xi=2$, respectively, while Fig. \ref{fig:ALP3} shows the same for the FSRQ ALP-corrected sources (all assume the conservative value $\sigma(\tau_{EBL})=0.2\tau_{EBL}$ for the EBL correction uncertainty).
For BL Lac objects, the effects of a lower cosmic opacity allowed by the photon-ALP mixing is apparent in softening the spectra at high $z$, when compared to the plots in Fig. \ref{fig:true}. The larger $\xi$ value in Fig. \ref{fig:ALP2} also produces a slightly larger increase of $\Gamma_{\rm em}$ with redshift.

The quantitative details of these new fits are reported in Tables \ref{tab:table4} and \ref{tab:tableIBL} for the HBL and IBL objects, and Table \ref{tab:table5} for FSRQs. The most significant effect is to increase the P-value probability of no evolution with redshift for the $\Gamma_{\rm em}$ of HBLs from $\sim 10\%$ to $\sim 50\%$, meaning that no correlation is at all present.
The results for the IBL and FSRQ classes are not significant due to the poor statistics.
Again, the $\chi^2$ and F-test do not offer any better sensitivity than the correlation tests. 

Altogether, the results of our analysis can be summarized with the statement that, so far, an effect at $\sim1.6 \ \sigma$ may be taken to indicate that some modifications of the conventional photon propagation process might be invoked, e.g. with the inclusion of the photon-ALP mixing, as discussed in \cite{2020MNRAS.493.1553G}. 
However, considering the far-reaching implications of such modifications, our conclusion is that standard-model physics ruling the photon-photon interaction process cannot so-far be rejected, in waiting for future much deeper TeV data to offer stronger conclusions. 

Note finally that, because of the saturation effect in the ALP correction mentioned in Sect. \ref{physical}, we cannot derive any significant constraints but only indications on the combined ALP parameter $\xi$ based on the present data.

%

\section{Conclusions}
\label{conclusions}
    
We have collected virtually all available data on the VHE spectra of blazars, including BL Lacs (both high- and intermediate-peaked, HBLs and IBLs, for a total of 68 spectra) and FSRQs (10 spectra).

We have re-analyzed all these data in terms of simple power-law spectral functions, whose best-fits clearly reveal the global effect of an increasing pair-production opacity with redshift due to interaction with EBL photons. 
After correcting all the individual spectral data for a conventional EBL extinction (including an up to ${\delta \tau}/{\tau}=0.2$ relative uncertainty in the EBL absorption corrections), the new spectral fits show indications for a residual anti-correlation of the intrinsic spectral indices at the source, $\Gamma_{\rm em}$, with redshift for both the IBL and the HBL samples, that were separately analyzed to avoid systematic effect induced by the blazar {\it spectral sequence} \citep[e.g.][]{1998MNRAS.299..433F}. As discussed in Sect. \ref{EBLcorrection}, whenever confirmed, this might eventually require some modification of the conventional photon propagation process.

The statistical significance of this effect is, however, limited to somewhat less than $2 \ \sigma$.
The few FSRQs available up to redshift $z\sim1$ do not add significant information, due to their limited statistics and large uncertainties in the spectral fits.

Motivated by these results and by previously published analyses, we have investigated the potential effects of introducing ALP-to-photon mixing, systematically lowering the cosmic photon-photon opacity. This is indeed a test offered by VHE astrophysics to validate (or not) an important physical effect potentially expected at the frontier of the standard model of fundamental interactions.
Under the assumption that such an ALP to photon conversion would operate in the presence of the EBL and of an intergalactic magnetic field \citep[see][]{2009MNRAS.394L..21D}, we have explored parameters of the photon-ALP system consistent with our data.
We find some weak indication in favour of such a photon-ALP mixing. Unfortunately, however, the presently available data do not set any constraints to the photon-ALP coupling $g_{a\gamma\gamma }$ and ALP mass $m_{\rm ALP}$, and standard model physics cannot be rejected with any confidence.

As we see, due to the limited statistics offered by the current VHE observations, our results cannot be taken at all as conclusive. VHE data by the forthcoming new generation of Cherenkov observatories, like LHAASO, CTA, ASTRI, SWGO, and others, will greatly improve these statistical constraints, particularly providing wider VHE spectral coverage for blazars on a larger redshift interval, with better statistics.

\section*{Acknowledgments}
We warmly thank Luca Foffano for insights on the {\it Fermi} Observatory data and on data analysis.
We are grateful to Renato Falomo for comments on the blazar redshift measurements.
Nijil Mankuzhiyil offered advice on MAGIC measurements for the source 1ES0033+595.
G.G. acknowledges a contribution from the grant ASI-INAF 2015-023-R.1.

The referee, Floyd Stecker, helped in significantly improving the paper, by pointing out an inaccuracy in an earlier paper version.
 
\section*{Data availability}
 All the relevant data and results of our analysis are incorporated into the article, together with a complete account of the bibliographic information about the sources where these data have been obtained.


\label{lastpage}
\end{document}